\documentclass[structabstract]{aa}
\usepackage{graphicx}
\usepackage{txfonts}
\usepackage{natbib}
\usepackage{longtable}
\usepackage{array}

\begin{document}

\title{\bf Twelve type II-P supernovae seen with the eyes of {\it Spitzer}}

\author{
T. Szalai
\and
J. Vink\'o
}

\institute{
Department of Optics and Quantum Electronics, University of
Szeged, D\'om t\'er 9., Szeged H-6720, Hungary\\
\email{[szaszi;vinko]@titan.physx.u-szeged.hu}
}

\date{Received ..; accepted ..}

\abstract{}{}{}{}{}

\abstract
{Core-collapse supernovae (CC SNe), especially those of type II-plateau (II-P), are thought to be important
contributors to cosmic dust production. The most obvious indicator of newly-formed and/or
pre-existing dust is the time-dependent mid-infrared (MIR) excess coming from the environment of SNe.
In the past few years several CC SNe were monitored by the Spitzer Space Telescope in the nebular phase,
hundreds of days after explosion. On the other hand, only a few of these objects have been analyzed
and published to date.}
{Our goal was to collect publicly available, previously unpublished measurements on type II-P (or peculiar IIP)
SNe from the Spitzer database. The most important aspect was to find SNe observed with the Infrared
Array Camera (IRAC) on at least two epochs. The temporal changes of the observed fluxes may be
indicative of the underlying supernova, while spectral energy distribution (SED) fitting to the fluxes in different IRAC channels may
reveal the physical parameters of the mid-IR radiation, which is presumably caused by warm dust.}
{The IRS spectra were extracted and calibrated with SPICE, while photometric SEDs were assembled using IRAF and MOPEX. Calculated SEDs from 
observed fluxes were fit with simple dust models to obtain basic information on the dust presumed as the source of MIR radiation.}
{We found twelve SNe that satisfied the criterion above, observed at late-time epochs (typically after +300
days). In three cases we could not identify any point source at the SN
position on late-time IRAC images. We found two SNe, 2005ad and 2005af, which likely have newly formed dust in their
environment, while in the other seven cases the observed MIR flux may originate from pre-existing
circumstellar or interstellar dust. Our results support the previous observational conclusions that warm new
dust in the environment of SNe contributes only marginally to the cosmic dust content.}
{}

\keywords{supernovae: general -- supernovae: individual -- dust, extinction -- infrared: stars}

\maketitle

\section{Introduction}\label{intro}

According to theoretical and observational studies during the past four decades, dust grains may be relevant components of the circumstellar matter (CSM) in 
the environment of core-collapse supernovae (CC SNe), which are the last stages of the life cycles of massive (M $\geq$ 8 M$_{\odot}$) stars. However, while the possibility 
of dust formation around CC SNe was first raised more than 40 years ago \citep{Cernuschi67,Hoyle70}, there are still many open questions in this topic.
In the local Universe, intense mass-loss processes of asymptotic giant branch (AGB) stars are thought to be the primary sources of dust. At the same time, the large 
observable amount of dust in high-redshift galaxies is hard to explain with AGB stars, especially because their formational timescale is longer than the age of these distant 
galaxies \citep[there are some detailed reviews about this topic, see e. g.][]{Dwek11,Gall11}. 

This is the point where CC SNe become relevant as major sources of dust in the early Universe. The lifetimes of their progenitors are usually shorter than $\sim$1 Gyr 
\citep[there are dust-rich high-z galaxies with this age, see e. g.][]{Maiolino04,Dwek07}. The classical models of dust production rates of CC SNe 
\citep{Kozasa89,Todini01,Nozawa03,Morgan03} all support this theory, predicting very similar masses (0.1-1 M$_\odot$) of newly formed dust. In more recent 
studies \citet{Bianchi07,Nozawa07,Kozasa09,Silvia10} also presented similar results. 

An important conclusion of the latter papers is that the amount and the basic 
properties of the dust formed in the ejecta depend strongly on the type of the progenitor, especially the thickness of its outer H and/or He layers. As 
\citet{Kozasa09} 
showed, a thicker outer envelope results in a lower expansion velocity and higher gas density in the deeper layers that contain the condensable elements (C, O, Mg, Al, Si). 
Therefore, larger ($\geq$0.03 $\mu$m) grains are able to form. This in turn means a larger initial dust mass and also a higher survival rate of the grains. This rate is also increased by 
the delayed arrival of the reverse shock at the dust-forming region (because of the lower gas density in the shocked region).
Based on these theoretical expectations, type II-P SNe are the best candidates for dust formation among CC SNe. Calculations of dust production efficiencies 
\citep{Gall11} also show that the mass range of 8-20 M$_{\odot}$ (which is thought to be the typical progenitor mass range of II-P SNe) is the most important for all SNe.

However, grain formation in the expanding and cooling ejecta is not the only way to find dust around CC SNe. If there is a relatively dense CSM around the progenitor 
(as it is in type IIn SNe), it has probably been formed via mass-loss processes from the progenitor. This pre-existing matter also contains dust grains and may also permit 
grain condensation interacting with the forward shock \citep[see more details e. g. in][]{Fox10,Fox11}. 

Although the results of different models seem to be concordant, observations of CC SNe in the local Universe do not support the prediction of the intensive 
dust production in these objects. Optical and mid-infrared analysis of the famous SN 1987A in the Large Magellanic Cloud (LMC) revealed $\sim$10$^{-4}$ M$_\odot$ 
of newly formed dust in the ejecta a few hundred days after explosion \citep[see e. g.][]{Danziger89,Danziger91,Lucy89,Lucy91,Roche93,Wooden93}, which is 
orders of magnitude lower than the value predicted by theoretical studies. Similar results were found in other CC SNe studied in the past two decades based primarily on the 
mid-IR data of the {\it Spitzer Space Telescope}, hereafter {\it Spitzer}. None of these objects were unambigously found to have more than 0.01 M$_\odot$ dust in their 
environment (see Section~\ref{conc} for more details). 

This controversy has been tried to be resolved in different ways. There is a widespread agreement that theoretical models are not perfect, and there are both micro- and 
macrophysical effects connected with dust formation that are not understood \citep[see e. g.][]{Cherchneff10,Fallest11}. Circumstellar dust may form optically thick 
clumps \citep{Lucy89}, resulting in a dust mass higher by an order of magnitude in numerical calculations \citep{Sugerman06,Ercolano07,Fabbri11}, but it is still not a satisfying 
explanation.
There is also a possibility that the missing dust mass is hidden in the form of cold ($\lesssim$ 50 K) grains, whose thermal radiation is mostly invisible for {\it Spitzer}'s 
instruments. This cold dust should stay mainly in the SN ejecta several years after explosion, in the so-called transitional phase 
\citep{Sugerman12,Tanaka12,Temim12a,Temim12b}, or in older supernova remnants (SNRs). Some of the 
nearby SNRs were observed at mid-IR, far-IR, and submillimeter (sub{\bf -}mm) wavelengths, but the inferred dust masses vary on a large scale (from 10$^{-4}$ to 1 M$_{\odot}$), 
sometimes even for the same object \citep[see the review of][]{Gall11}. Recently, observations from the 
{\it Herschel Space Observatory} (hereafter {\it Herschel}) are expected to give reassuring answers. The published results suggest relatively large dust masses: 
0.08 M$_{\odot}$ in Cas A \citep{Barlow10}, 0.1-10 M$_{\odot}$ in N49 in the LMC \citep{Otsuka10}, and 0.4-0.7 M$_{\odot}$ in SN 1987A \citep{Matsuura11}; although new 
measurements on SN~1987A made with the Atacama Pathfinder EXperiment (APEX) at sub-mm wavelengths do not support the presence of such a large amount of dust 
\citep{Lakicevic12}.

There are also some other ideas, but they are not supported by any convincing observational evidence. The observed low dust production rates of CC SNe could be 
compensated with a top-heavy initial mass function (IMF), which might be a characteristic of some early galaxies \citep[e. g.][]{Bromm02,Tumlinson06,Michalowski10}. It is 
also possible that pair-instability supernovae (PISNe), the assumed fate of stars with progenitor masses of 140-260 M$_{\odot}$ \citep{Heger03}, could be 
efficient dust producers in the first galaxies \citep{Nozawa03,Gall11}, but there are still no strong evidence for the existence of these objects. 
Some authors argue that depending on star-formation history, AGB stars may be significant dust sources in galaxies younger than 1 Gyr \citep{Valiante09,Dwek11}.
There is another interesting idea about possible grain condensation in quasar winds \citep{Elvis02}. Beyond these ideas, more calculations suggest significant grain 
growth in the ISM, which may be an acceptable explanation of both the low dust production rates of SNe and the high dust content of distant galaxies 
\citep{Draine03,Draine09,Michalowski10,Mattsson12a,Mattsson12b}.

To get closer to the mystery of the origin of cosmic dust, it is necessary to have a large amount of high-quality observational data. Up to now, there are only 20-25 SNe 
for which we have quantitative pieces of information about their dust content; see e. g. \citet{Fox11}, \citet{Gall11}, \citet{Szalai11}, and Section~\ref{conc} for details. 
Regarding  the type II-P SNe, there 
are less than ten objects with detailed analyses, and we know quite a few examples without detectable dust content. \citet{Fox11} published the results of a 
survey to 
detect MIR excess (the most obvious evidence of warm dust) in the environment of type IIn SNe. They found late-time emission in ten objects from their sample of 69 SNe and 
concluded that the IIn subclass represents the largest part of dust-containing SNe (it should be noted that the authors carried out this project during the 
warm mission of {\it Spitzer}, so they were able to use only the 3.6 and 4.5 $\mu$m channels supplemented by optical and near-IR photometry and spectroscopy).
 
Motivated partly by the work of \citet{Fox11}, our goal is to perform such analyses for type II-P SNe that are thought to be the best candidates for producing 
newly formed dust. 
We collect, reduce, and analyze unpublished measurements on CC SNe belonging to the II-P subclass, found in the public 
{\it Spitzer} database. In Section~\ref{obs} we show the steps of data reduction and present the extracted photometric and spectroscopic data. We present observed spectral 
energy distributions (SEDs) of the studied SNe, and compare them with different dust models (Section~\ref{anal}).
We note that although the properties of the cold ISM are better constrained by far-IR and/or sub-mm data, the low angular resolution of the available instruments prevent 
measurements of extragalactic SNe. The {\it Spitzer} data we applied in this paper probe only the warm ($T \sim 200$ - 1000 K) dust in the environment of extragalactic SNe.
Thus, our study focuses on the presence/absence of warm dust and its physical properties. We discuss the results of the SED fitting in Section \ref{models}.
Finally, we summarize our results and conlusions, and give a brief discussion on the role of type II-P SNe in cosmic dust production.

\section{Observations and data reduction}\label{obs}

We queried the public {\it Spitzer} database to find type II-P SNe observed with the  Infrared Array Camera (IRAC) on at least two epochs. 
We found 12 SNe satisfying this criterion. We also collected observations obtained by {\it Spitzer}'s
Multiband Imaging Spectrometer (MIPS) and Infrared Spectrograph (IRS) instruments (not only the spectra but also the broad-band photometric data recorded in 
peak-up imaging mode, PUI).
In addition, we also checked the literature for any other observational data obtained around the same epoch as the {\it Spitzer} measurements.

\subsection{Supernova sample observed by {\it Spitzer}}\label{sn}

\begin{table*}
\caption{\label{tab:data} Basic data of the supernovae}
\centering
\newcommand\T{\rule{0pt}{3.1ex}}
\newcommand\B{\rule[-1.7ex]{0pt}{0pt}}
\begin{tabular}{llllllccl}
\hline
\hline
Name & RA & DEC & Galaxy & Date of & MJD $-$ & D (Mpc) & E(B$-$V) & References \T \\
 & (J2000) & (J2000) & & explosion & 2\,450\,000 & & (mag) & \B \\
\hline
SN~2003J & 12:10:57.7 & +50:28:31.8 & NGC~4157 & 2003-01-11$^{\dagger}$  & 2651$^{\dagger}$  & 14.7 & 0.021$^{\ddagger}$ & 1, 2 \T \\
SN~2003hn & 03:44:36.1 & $-$44:37:49.0 & NGC~1448 & 2003-08-07 & 2859 & 18.1 & 0.19  & 3, 4, 5 \\
SN~2003ie & 12:03:18.1 & +44:31:36.8 & NGC~4051 & 2003-08-16 & 2868 & 15.5 & 0.013$^{\ddagger}$ & 1, 6, 7, 8 \\
SN~2004A & 16:43:01.9 & +36:50:12.5 & NGC~6207 & 2004-01-06 & 3011 & 20.3 & 0.06  & 9 \\
SN~2005ad & 02:28:29.4 & $-$01:08:20.0 & NGC~941  & 2005-02-06$^{\dagger}$  & 3407$^{\dagger}$ & 20.8 & 0.035$^{\ddagger}$ & 1, 10 \\
SN~2005af & 13:04:44.1 & $-$49:33:59.8 & NGC~4945 & 2005-01-07 & 3379 & 3.9  & 0.183$^{\ddagger}$ & 1, 11 \\
SN~2005cs & 13:29:52.8 & +47:10:36.1 & NGC~5194 & 2005-06-28 & 3550 & 8.4  & 0.04 & 12, 13, 14, 15 \\
SN~2006bc & 07:21:16.5 & $-$68:59:57.3 & NGC~2397 & 2006-03-24 & 3819 & 14.7 & 0.205$^{\ddagger}$ & 1, 7, 16 \\
SN~2006bp & 11:53:55.7 & +52:21:09.4 & NGC~3953 & 2006-04-07 & 3833 & 17.5 & 0.400 & 17 \\
SN~2006my & 12:43:40.7 & +16:23:14.1 & NGC~4651 & 2006-07-26 & 3943 & 22.3 & 0.027$^{\ddagger}$ & 1, 18, 19, 20 \\
SN~2006ov & 12:21:55.3 & +04:29:16.7 & NGC~4303 & 2006-08-16 & 3964 & 12.6 & 0.022$^{\ddagger}$ & 1, 7, 19 \\
SN~2007oc & 22:56:41.8 & $-$36:46:22.3 & NGC~7418A & 2007-11-03$^{\dagger}$ & 4408$^{\dagger}$ & 28.0 & 0.023$^{\ddagger}$ & 1, 21  \B \\
\hline
\end{tabular}
\tablebib{
(1)~\citet{Schlegel98}; (2) \citet{Tonry01}; (3) \citet{Jones09}; (4) \citet{Krisciunas09}; (5) \citet{Olivares10}; (6) \citet{Pierce88};
(7) \citet{Smartt09}; (8) \citet{Harutyunyan08};
(9) \citet{Hendry06}; (10) \citet{Li11}; (11) \citet{Kotak06}; (12) \citet{Pastorello06}; (13) \citet{Baron07}; (14) \citet{Pastorello09}; (15) \citet{Vinko12}; 
(16) \citet{Gallagher12}; (17) \citet{Dessart08}; (18) \citet{Solanes02}; (19) \citet{Li07}; (20) \citet{Maguire10}; (21) \citet{GildePaz07}.
}
\tablefoot{
$^{\dagger}$ From date of discovery (date of explosion is unknown).\\
$^{\ddagger}$ Galactic reddening.
}
\end{table*}

{\bf SN~2003J} was discovered in NGC~4157 on 2003 January 11.3 UT (dates are in UT thereafter) by \citet{Kushida03} ($\sim$ 16.7 mag, unfiltered). 
\citet{Ayani03} classified 2003J as a normal type-II SN. Using the infrared map of \citet{Schlegel98}, the calculated value of galactic interstellar reddening 
is $E(B-V)_{gal}$ = 0.021 mag. The estimated distance ($D$) to the host galaxy (based on surface brightness fluctuations) is 14.7 Mpc \citep{Tonry01}.

{\bf SN~2003hn} was discovered in NGC~1448 on 2003 August 25.7 by \citet{Evans03} ($V \sim$ 14.4 mag). Preliminary spectroscopic analysis by \citet{Salvo03} showed that 
2003hn was a type II-P supernova. Detailed analysis based on optical and near-infrared photometry and optical spectroscopy was carried out by \citet{Krisciunas09}, while 
\citet{Jones09} and \citet{Olivares10} also analyzed the data of SN~2003hn and of other type II-P SNe to obtain precise distances to the host galaxies. 
We adopt 18.1 Mpc as the distance of the SN \citep{Krisciunas09} and  $t_0$ = JD 2\,452\,859 as the moment of explosion \citep{Jones09}, while we use 
$E(B-V)_{total}$ = 0.19 mag as the estimated reddening of the host galaxy \citep{Olivares10}.

{\bf SN~2003ie} was discovered in NGC~4051 on 2003 September 19.8 \citep{Arbour03} with an apparent brightness of $\sim$ 15.0 mag (unfiltered). \citet{Benetti03} classified 
it as a type-II SN, while \citet{Harutyunyan08} found similarities between some spectral features of SN~2003ie and the peculiar type II-P SNe 1998A and 1987A. Based on 
data on both the SN and its host galaxy (distance, metallicity, extinction) \citet{Smartt09} gave 25 M$_{\odot}$ as an upper limit for its progenitor mass. For our studies we 
adopt $D$ = 15.5 Mpc \citep{Pierce88,Smartt09}, $E(B-V)_{gal}$ = 0.013 mag \citep{Schlegel98}, and $t_0$ = 2\,452\,868 \citep{Harutyunyan08}.

{\bf SN~2004A} was discovered in NGC~6207 on 2004 January 9.8 by the Japanese amateur astronomer K. Itagaki \citep[see][]{Nakano04} with a brightness of $\sim$ 15.7 mag 
(unfiltered). A preliminary optical spectrum obtained by \citet{Kawakita04} indicated that it was a type-II SN. 
Based on their detailed optical photometric and spectroscopic analysis, 
\citet{Hendry06} found 2004A to be a classical type II-P SN with $D$ = 20.3 Mpc, $E(B-V)_{total}$ = 0.06 mag, and $t_0$ = 2\,453\,011. \citet{Maguire10} also published optical 
photometric and spectroscopic data from the photospheric phase.

{\bf SN~2005ad} was discovered in NGC~941 on 2005 February 6.4 by K. Itagaki on unfiltered CCD-images \citep[see][]{Nakano05} with a brightness of $\sim$ 17.4 mag.
The very blue continuum on early spectra \citep{Morrell05} indicated that 2005ad was discovered in a very early phase, but the exact value of $t_0$ is still unknown. 
Lacking detailed studies, classification of 2005ad is difficult, but some authors \citep{Smartt09,Li11} refer to it as a type II-P SN. We use the values of 
$D$ = 20.8 Mpc \citep{Li11} and $E(B-V)_{gal}$ = 0.06 mag \citep{Schlegel98}.

{\bf SN~2005af} appeared in the nearby galaxy NGC~4945 on 2005 February 8.2 with an apparent brightness of $V$ = 12.8 mag \citep{Jacques05}. Spectroscopic analysis of 
\citet{Filippenko05} showed that it was a normal type II-P event. \citet{Pereyra06} obtained optical polarimetry during the photospheric and the early 
nebular phase, while \citet{Kotak06} and \citet{Kotak08} reported mid-IR data from this period (see details below). From the latter study we adopt $D$ = 3.9 Mpc 
and $t_0$ = 2\,453\,379, and use $E(B-V)_{gal}$ = 0.183 mag for reddening \citep{Schlegel98}.

{\bf SN~2005cs} was discovered in the nearby, spectacular Whirlpool Galaxy (M51, NGC~5194) on 2005 June 28.9 by \citet{Kloehr05}. It was classified as a type-II SN by 
\citet{Modjaz05}. Because of its proximity, it was the target of many observing campaigns. Early-time optical photometry and spectroscopy were presented by 
\citet{Pastorello06} and \citet{Dessart08}, while \citet{Brown07} studied the SN using the {\it Swift} ultraviolet-optical and X-ray telescopes. SN~2005cs seems to belong to the
subtype of low-luminosity, $^{56}$Ni-poor, low-velocity SNe II-P, as shown also by long-term studies of \citet{Tsvetkov06} (optical photometry), \citet{Gnedin07} (R-band 
photometry and polarimetry), and \citet{Pastorello09} (optical and near-IR photometry, optical spectroscopy). Extensive studies of the first-year data and pre-explosion
images resulted in a progenitor mass of 7--13 M$_{\odot}$ \citep{Maund05,Li06,Takats06,Eldridge07}, while \citet{Utrobin08} determined a significantly higher value 
(17--19 M$_{\odot}$). In their more recent work \citet{Pastorello09} concluded that the progenitor mass should be in the range of 10--15 M$_{\odot}$. The distance to the 
host galaxy, determined via different methods, is scattered between 5.9--9.4 Mpc \citep{Feldmeier97,Pastorello06,Takats06,Dessart08}. Recently, \citet{Vinko12} determined 
its mean value of 8.4 Mpc by combining the data of SNe 2005cs and 2011dh in the same galaxy. We adopt this latter value for $D$ as well as $t_0$ = 2\,453\,550 
\citep{Pastorello06}, and $E(B-V)_{total}$ = 0.04 mag \citep{Baron07,Dessart08}.

{\bf SN~2006bc} was discovered in NGC~2397 by R. Martin \citep[see][]{Monard06} on 2006 March 24.6; its apparent brightness was 16.0 mag (on an unfiltered CCD-image). Early 
spectra clearly showed that it was a type-II SN \citep{Patat06}, but due to the lack of long-term studies it remains unsolved whether it is a II-P or a II-L SN 
\citep{Smartt09,Otsuka12}. While \citet{Smartt09} could not identify the progenitor of 2006bc on {\it HST} archive images, \citet{Otsuka12} also used {\it HST} to obtain 
late-time optical photometry. \citet{Brown09} also observed it with the {\it Swift} UVOT detector. Parallel to our work, \citet{Gallagher12} carried out a 
complete optical and infrared analysis (see details below). The distance to the host galaxy, $D$ = 14.7 Mpc, is only 
estimated from its kinematic motion \citep{Smartt09}. We also adopt $E(B-V)_{gal}$ = 0.205 mag \citep{Schlegel98}, and $t_0$ = 2\,453\,819 \citep{Gallagher12}.

{\bf SN~2006bp}, appearing in NGC~3953, was found by K. Itagaki \citep[see][]{Nakano06a} on 2006 April, 9.6 (16.7 mag, unfiltered). Early spectroscopic 
\citep{Quimby06,Quimby07,Dessart08} and photometric observations -- in optical \citep{Quimby07,Dessart08}, UV and X-ray \citep{Immler07,Dessart08,Brown09} -- showed that 
2006bp was a classical type II-P SN, and it was discovered only a few hours after shock breakout, which is the earliest discovery among type II-P SNe. Adopting the results of
\citet{Dessart08}, we used the data $D$ = 17.5 Mpc, $E(B-V)_{total}$ = 0.400 mag, and $t_0$ = 2\,453\,833.

{\bf SN~2006my} was discovered in NGC~4651 by K. Itagaki \citep[see][]{Nakano06b} on 2006 November 8.8 with an apparent unfiltered brightness of 15.3 mag, well past maximum 
light. Analyzing the first spectra after discovery, \citet{Stanishev06} classified it as a type II-P SN. \citet{Li07} and \citet{Maguire10} carried out optical photometric 
and spectroscopic observations both in the photospheric and in the early nebular phase. From optical spectropolarimetry, \citet{Chornock10} showed strong late-time 
asphericities in the 
inner core of 2006my. Extended studies, based on the analysis of pre- and post-explosion images of HST and high-resolution ground-based telescopes, concluded that the progenitor 
of the SN should have been a red supergiant with a mass between 7--13 M$_{\odot}$ \citep{Li07,Leonard08,Smartt09,Crockett11}. \citet{Solanes02} reported $D$ = 22.3 Mpc as the 
distance of the host galaxy; we used this value, as well as $t_0$ = 2\,453\,943 as the moment of explosion \citep{Maguire10} and $E(B-V)_{gal}$ = 0.027 mag as galactic reddening 
\citep{Schlegel98}.

{\bf SN~2006ov}, appearing in NGC~4303, has many similarities with 2006my. It was also found by K. Itagaki \citep[see][]{Nakano06c} on 2006 November 24.9. Its 
unfiltered apparent brightness was 14.9 mag at the moment of discovery. Similarly to 2006my, 2006ov was a type II-P SN discovered near the end of the 
plateau phase \citep{Blondin06,Li07}. It was also one of the three CC SNe in which \citet{Chornock10} found late-time asphericity in the inner ejecta. \citet{Li07} determined 
the mass of the progenitor as 15$^{+5}_{-3}$ M$_{\odot}$, but later works \citep{Smartt09,Crockett11} gave 10 M$_{\odot}$ as an upper limit. We adopt $D$ = 12.6 Mpc
\citep{Smartt09} and $E(B-V)_{gal}$ = 0.022 mag \citep{Schlegel98}, while we estimated $t_0$ = 2\,453\,964 based on the results of \citet{Li07}.

{\bf SN~2007oc} was discovered in NGC~7418A on 2007 November 3.1 by the Chilean Automatic Supernova Search program \citep{Pignata07}. Due to the lack of any published 
photometric or spectroscopic observations, the moment of explosion and its subtype remain unknown; it is mentioned in the ASIAGO Supernova Catalogue as a type II-P 
SN \citep{Barbon08}. We adopted $D$ = 28.0 Mpc as the distance of the host galaxy \citep{GildePaz07} and $E(B-V)_{gal}$ = 0.023 mag \citep{Schlegel98} as the 
galactic reddening.

\subsection{Mid-infrared photometry with Spitzer}

\begin{figure*}
\centering
\includegraphics[width=5cm, height=3.54cm]{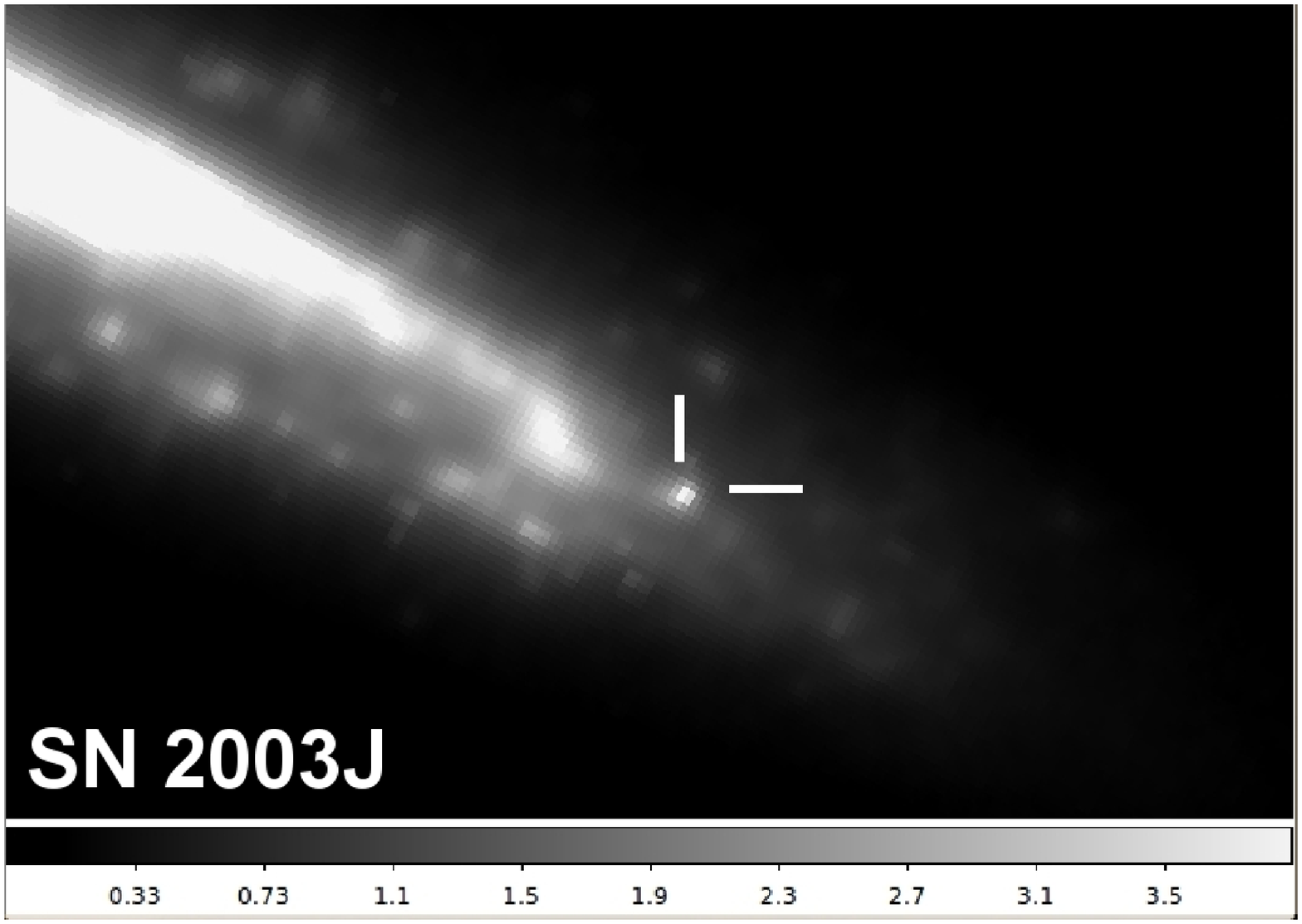} \vspace{3mm} \hspace{3mm}
\includegraphics[width=5cm, height=3.54cm]{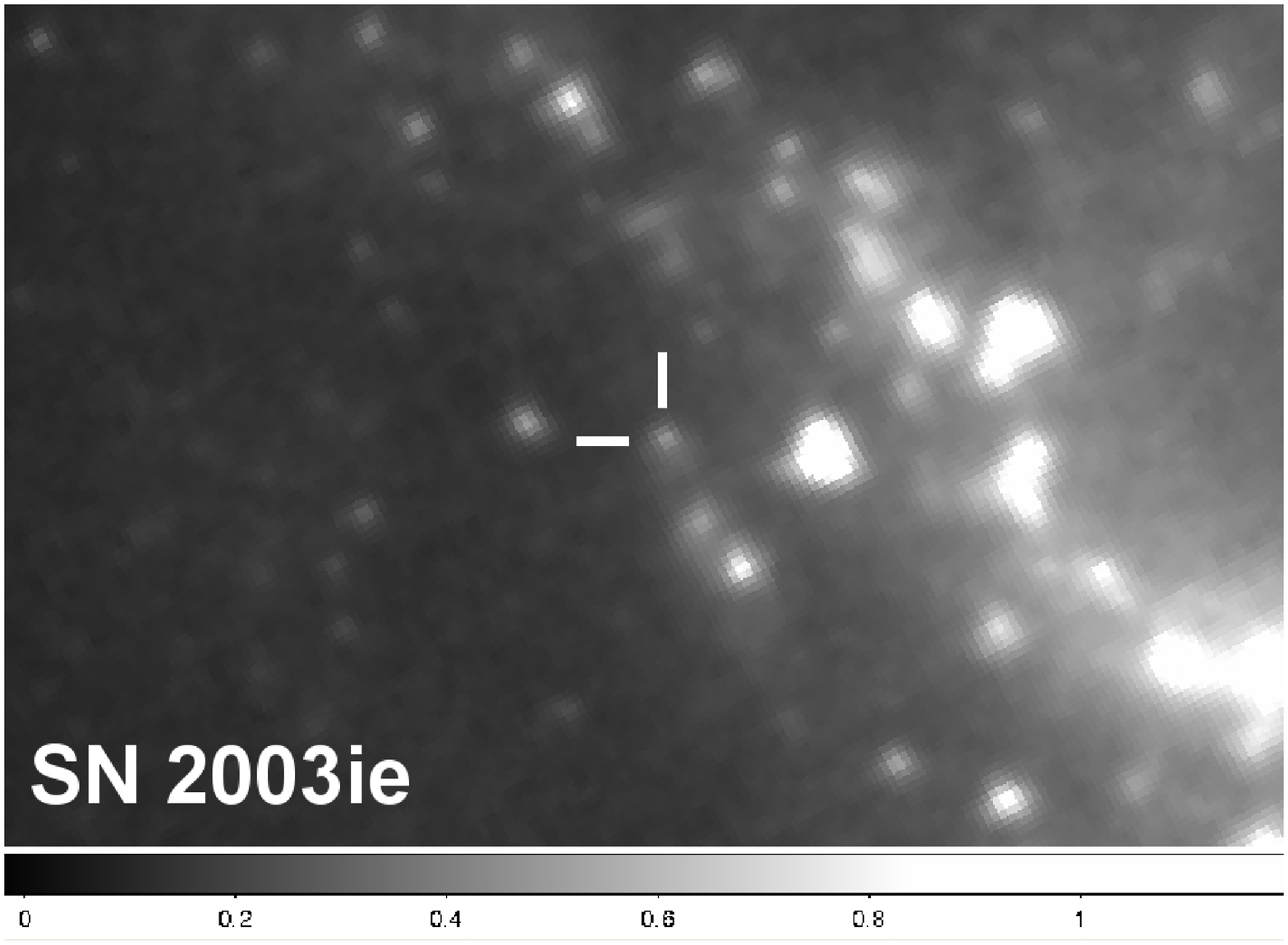} \hspace{3mm}
\includegraphics[width=5cm, height=3.54cm]{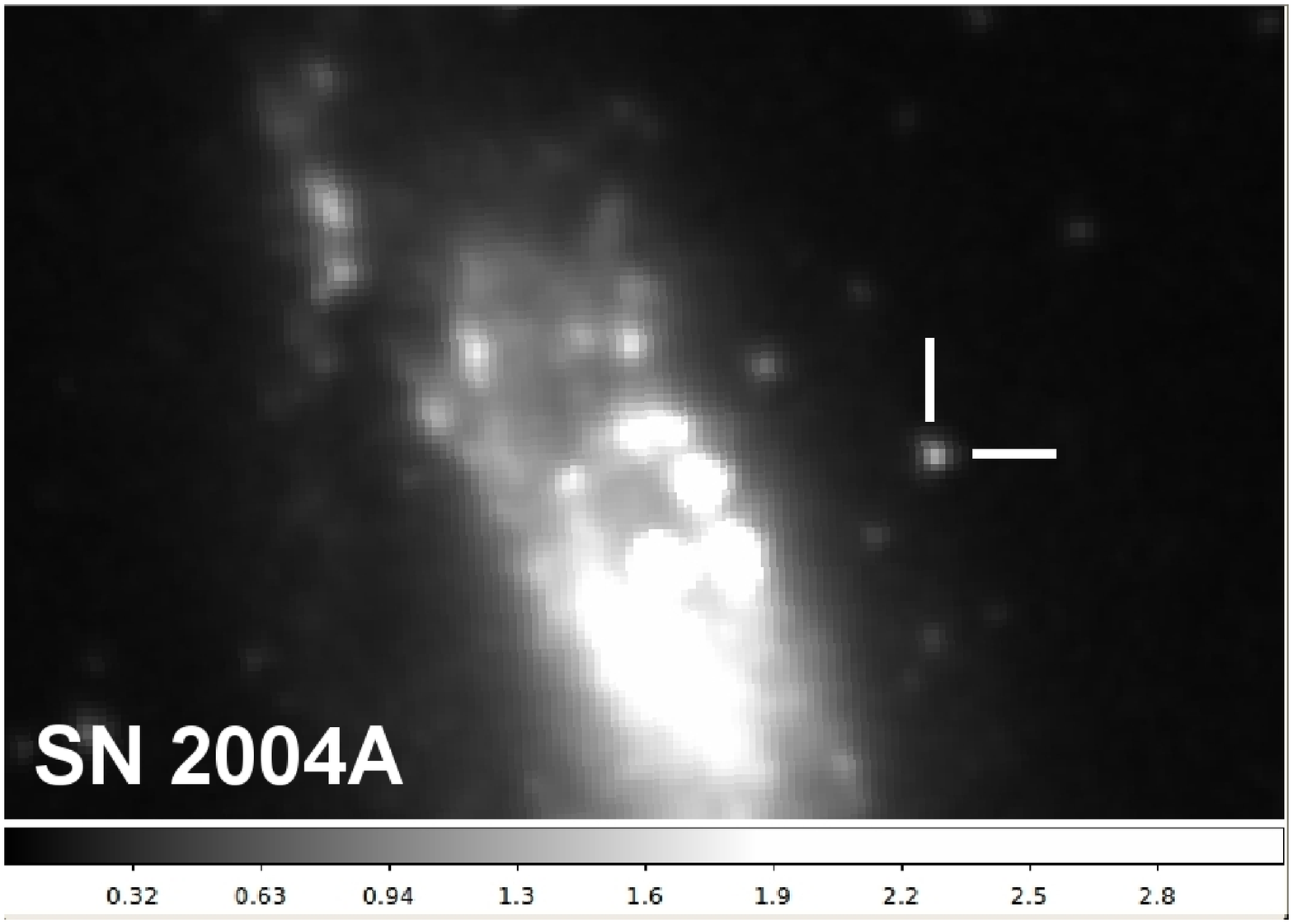} \vspace{3mm}
\includegraphics[width=5cm, height=3.54cm]{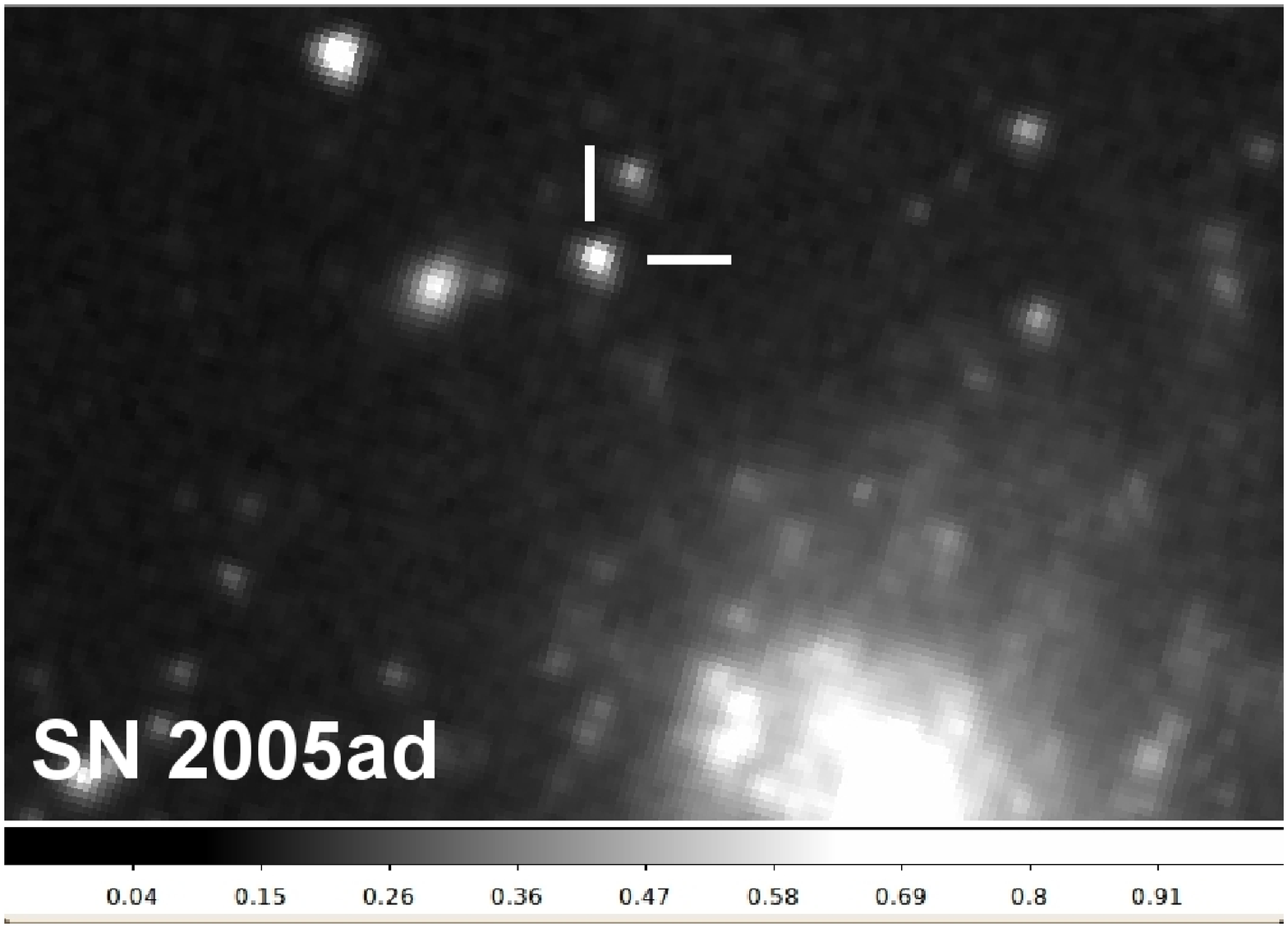} \hspace{3mm}
\includegraphics[width=5cm, height=3.54cm]{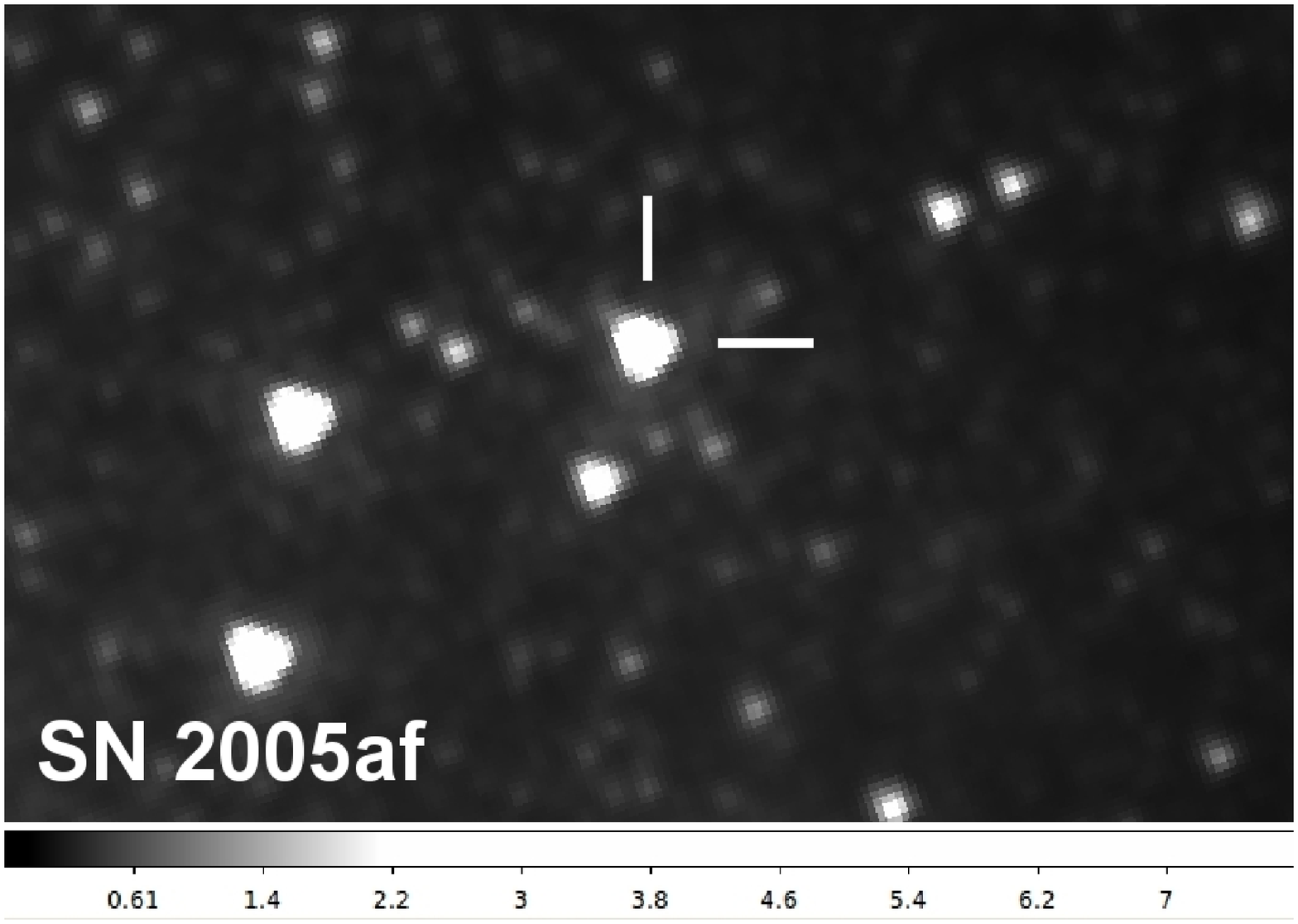} \hspace{3mm}
\includegraphics[width=5cm, height=3.54cm]{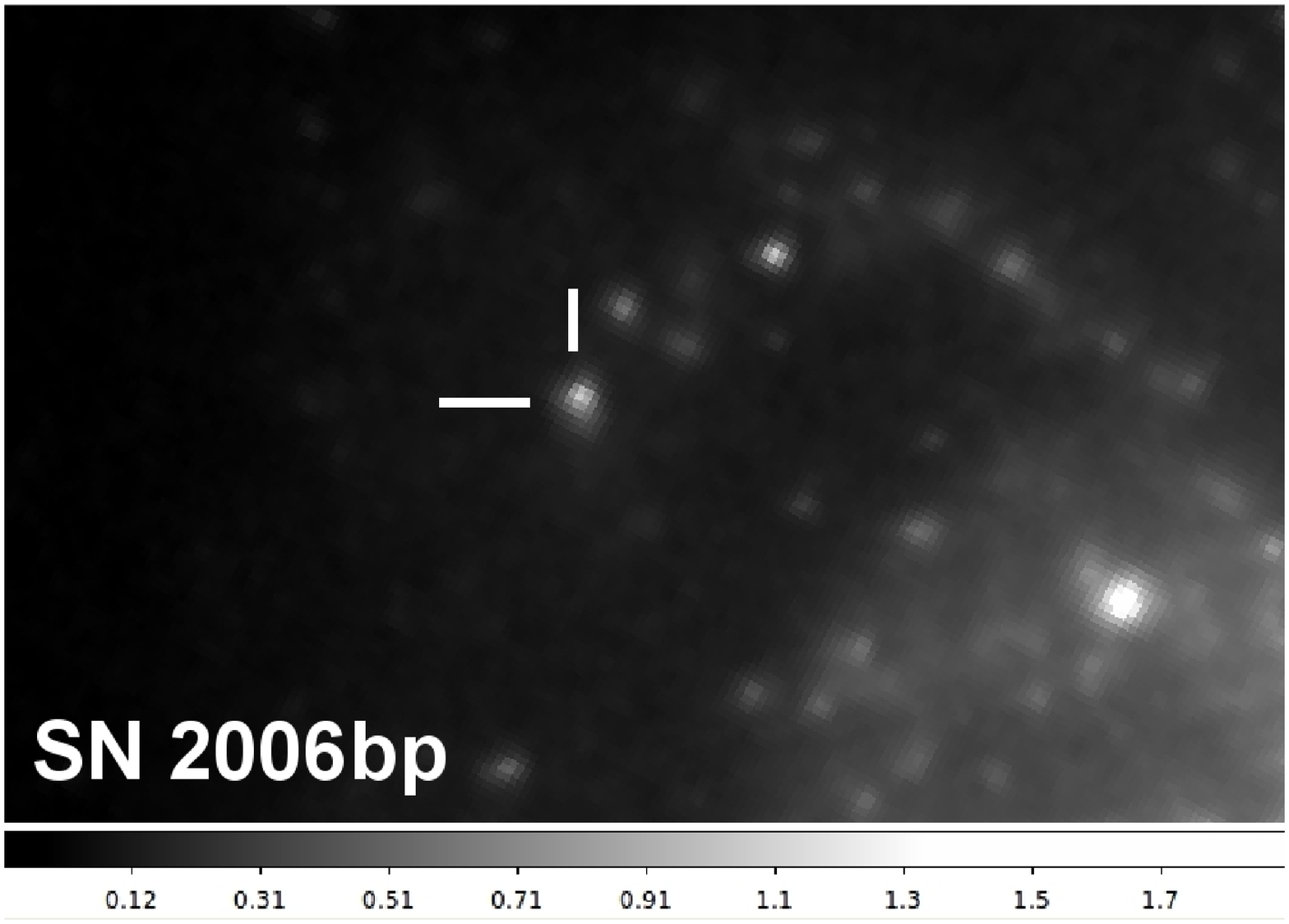} 
\includegraphics[width=5cm, height=3.54cm]{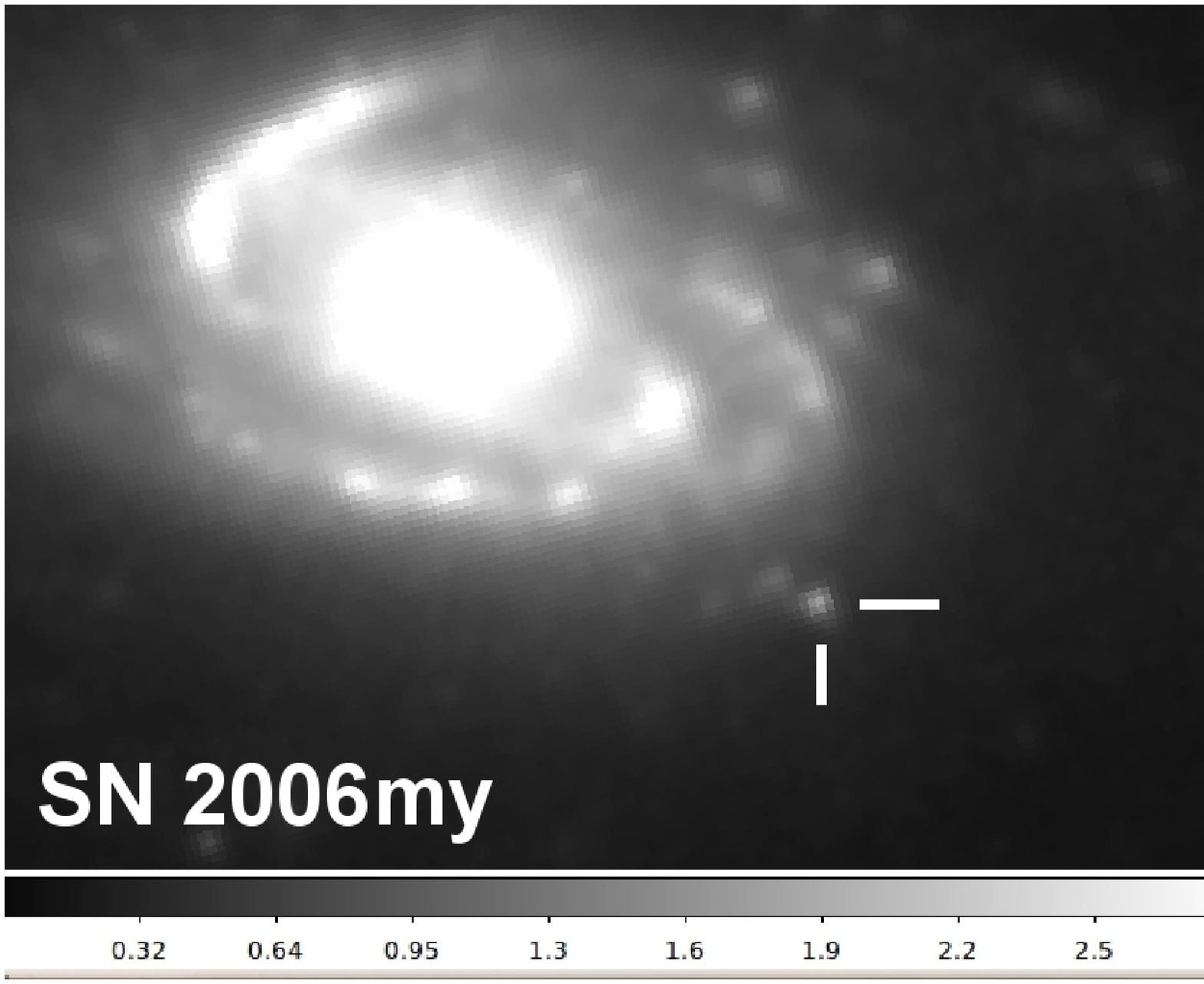} \hspace{3mm}
\includegraphics[width=5cm, height=3.54cm]{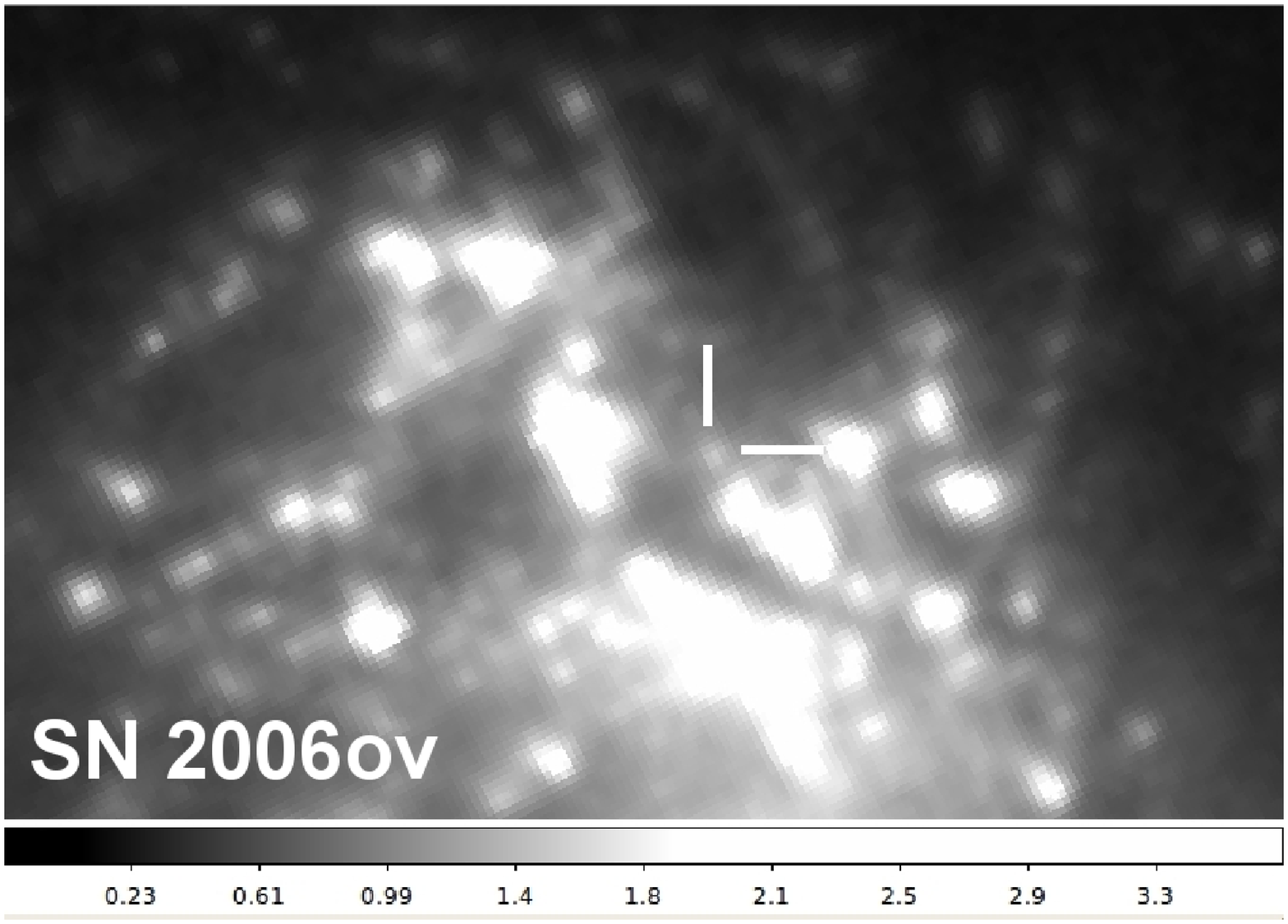} \hspace{3mm}
\includegraphics[width=5cm, height=3.54cm]{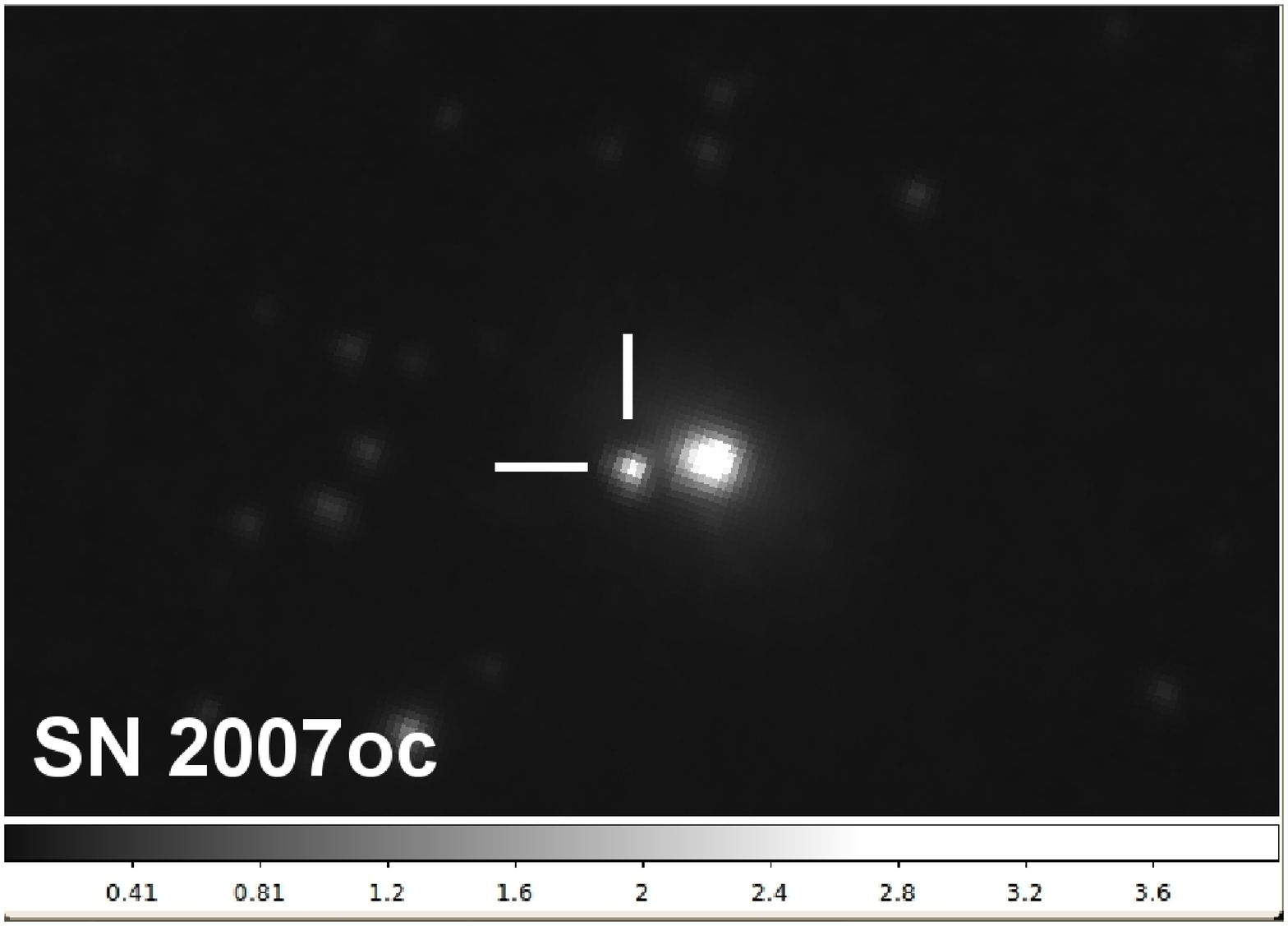}
\caption{Type II-P SNe on IRAC post-BCD 4.5 $\mu$m images. The FOV is 100\arcsec $\times$ 60\arcsec. North is up, and east is to the left.}
\label{fig:SNe}
\end{figure*}

We collected and analyzed all available IRAC post-basic calibrated data (PBCD) on the 12 CC SNe listed above. The scale of those images is 0.6$\arcsec$/pixel.
In some cases, a clear identification of SN was not easy because of the faintness of the target in the vicinity of bright regions. 
We tried to apply the image-subtraction technique used succesfully in some cases to analyze {\it Spitzer} data of SNe \citep{Meikle06,Meikle07,Kotak09}. We found 
pre-explosion images of the host galaxies for only two SNe (2005cs and 2006ov), and their quality was too poor for reliable image subtraction. We also generated some 
subtracted images using the latest frames as templates, but their applicability was also very limited.
Finally, we were able to unambiguously identify a MIR point source in nine cases at the SN position, while in the cases of three SNe (2003hn, 2005cs, and 2006bc) we did 
not find any sign of a point source, either on the original images or on the subtracted image (however, as we mentioned before, there are some other recent results about 
SN~2006bc, see details in Section~\ref{conc}). For the nine detected SNe (Fig. \ref{fig:SNe}) we carried out a detailed photometric analysis.  

Originally, we planned to apply PSF-photometry using the {\texttt DAOPHOT} package of IRAF\footnote{IRAF is distributed by the National Optical Astronomy 
Observatories, which are operated by the Association of Universities for Research in Astronomy, Inc., under cooperative 
agreement with the National Science Foundation.} to obtain more accurate results. Unfortunately, the bright background flux of the host galaxies, sometimes showing rapid 
spatial variations, and the lack of relatively bright point sources on the frames prevented us from applying this. 

Therefore we carried out simple aperture photometry on the PBCD frames with the {\texttt PHOT} task of the IRAF software package. This was possible beacuse the studied SNe -- 
except of the three mentioned above -- were clearly identifiable on IRAC images (see Fig. \ref{fig:SNe}). We generally used an aperture radius of 3.6\arcsec 
and a background annulus from 3.6\arcsec to 8.4\arcsec 
(3-3-7 configuration in native $\sim$1.2\arcsec IRAC pixels), applied aperture corrections of 1.124, 1.127, 1.143, and 1.234 for the four IRAC channels (3.6, 4.5, 5.8, and 8.0 $\mu$m) as given in 
the IRAC Data Handbook \citep{Reach06}. We chose a 2-2-6 configuration for SNe 2003J and 2007oc (aperture corrections: 1.213, 1.234, 1.379, and 1.584), and a 5-12-20 
configuration on the images of the bright, nearby SN~2005af (aperture corrections: 1.049, 1.050, 1.058, and 1.068).

There is one object, 2005af, for which MIR data were partly analyzed and published before: 
\citet{Kotak06} presented IRS spectra and also IRAC and PUI fluxes up to +433 days, while in \citet{Kotak08} there is a figure showing an IRS spectrum and IRAC measurements 
carried out at +571 days. To check the reliability of our reduction technique we also reduced these data and found a good agreement with the results of \citet{Kotak06} and 
\citet{Kotak08}. We also completed this dataset with the fluxes determined from the latest measurements found in the {\it Spitzer} database.

We also carried out aperture photometry using IRAF on late-time MIPS 24.0 $\mu$m PBCD frames (image scale 2.45\arcsec/pixel) 
for SNe 2003J, 2003ie, 2005af, 2006bp, and 2006my. We used an aperture radius of 3.5\arcsec and a background annulus from 6\arcsec to 8\arcsec, except for SN~2005af, for which we
found the 5-5-12 configuration to be the best choice. 
For SNe 2005af and 2006my, the MIR point sources were also identifiable on available IRS PUI images. Their fluxes
were measured using the MOsaicker and Point source EXtractor ({\it MOPEX}\footnote{http://ssc.spitzer.caltech.edu/dataanalysistools/tools/mopex/}) software.
All measured fluxes are listed in Table \ref{phot}.

\subsection{Mid-infrared spectroscopy with IRS}

We also downloaded the available IRS spectra of the SNe from the {\it Spitzer} database. Four spectra of SN~2005af 
-- the first three have already been published by \citet{Kotak06} and \citet{Kotak08} --, two spectra of SN~2004A, and a single spectrum of SNe 2003ie, 2005ad, 2005cs, 2006bp, 
and 2006ov were collected this way. 

\begin{table*}
\caption{\label{tab:spec} IRS observations of the SNe}
\centering
\newcommand\T{\rule{0pt}{3.1ex}}
\newcommand\B{\rule[-1.7ex]{0pt}{0pt}}
\begin{tabular}{llcccl}
\hline
\hline
Object & UT Date & MJD $-$ & $t-t_{expl}$ & ID & Proposal ID (PID) \T \\
 & & 2\,450\,000 & (days) & & \B \\
\hline
SN~2003ie & 2005-01-14 & 3385 & 517 & r10557696 & 3248 (Meikle et al.) \T \\
SN~2004A & 2004-08-27 & 3244 & 233 & r10557952 & 3248 (Meikle et al.) \\
 & 2005-02-16 & 3417 & 406 & r10558208 & 3248 (Meikle et al.) \\
SN~2005ad & 2005-09-08 & 3621 & 215$^{\dagger}$ & r14466304 & 20256 (Meikle et al.) \\
SN~2005af & 2005-03-17 & 3446 & 67 & r13413376 & 237 (Van Dyk et al.) \\
 & 2005-08-11 & 3593 & 214 & r14468096 & 20256 (Meikle et al.) \\
 & 2006-08-03 & 3950 & 571 & r17969664 & 30292 (Meikle et al.) \\
 & 2007-03-25 & 4185 & 806 & r17969920 & 30292 (Meikle et al.) \\
SN~2006bp & 2008-01-18 & 4483 & 648 & r23111936 & 40619 (Kotak et al.) \B \\
\hline
\end{tabular}
\tablefoot{
$^{\dagger}$ From date of discovery.
}
\end{table*}

We analyzed the PBCD-frames of the spectroscopic data with the SPitzer IRS Custom Extraction software 
({\it SPICE}\footnote{http://ssc.spitzer.caltech.edu/dataanalysistools/tools/spice/}). Sky subtraction and bad pixel removal were performed using two exposures that contained
the spectrum at different locations, which were then subtracted them from each other. Order extraction, wavelength- and flux-calibration were performed by applying built-in templates 
within {\it SPICE}. Finally, the spectra from the first, second and third orders were combined into a single spectrum with the overlapping edges averaged. In a few cases the sky 
near the order edges was oversubtracted because of some excess fluxes close to the source, which resulted in spurious negative flux values in the extracted spectrum. These 
were filtered out using the fluxes in the same wavelength region that were extracted from the adjacent orders. 

\begin{figure}
\centering
\resizebox{\hsize}{!}{\includegraphics{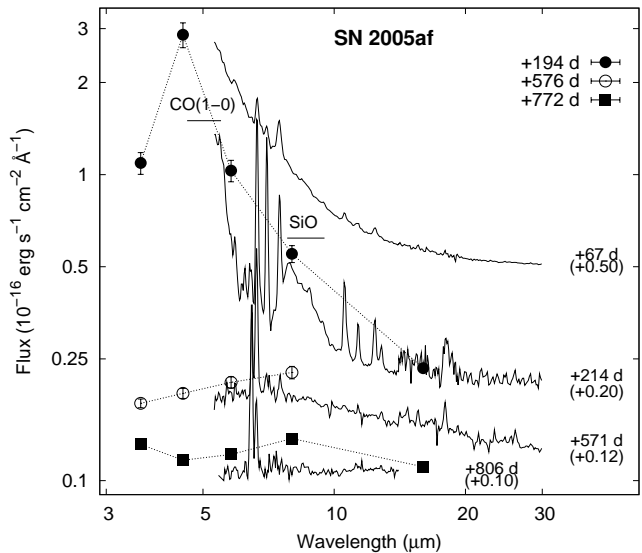}}
\caption{IRS spectra of SN~2005af. The spectra at different epochs are vertically shifted relative to each other by the amount indicated in the labels next 
to the spectra.}
\label{fig:2005af_irs}
\end{figure}

\begin{figure*}
\begin{center}
\includegraphics[width=7.5cm, height=5.25cm]{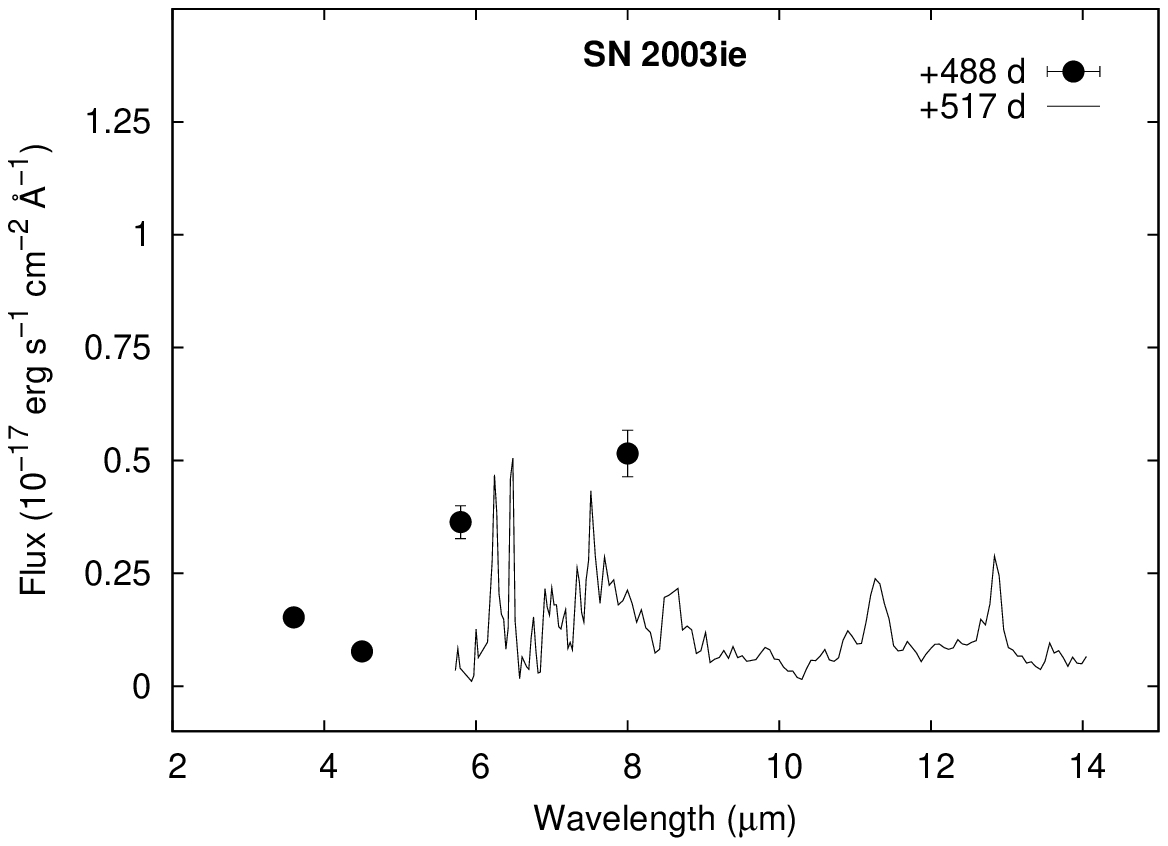}
\includegraphics[width=7.5cm, height=5.25cm]{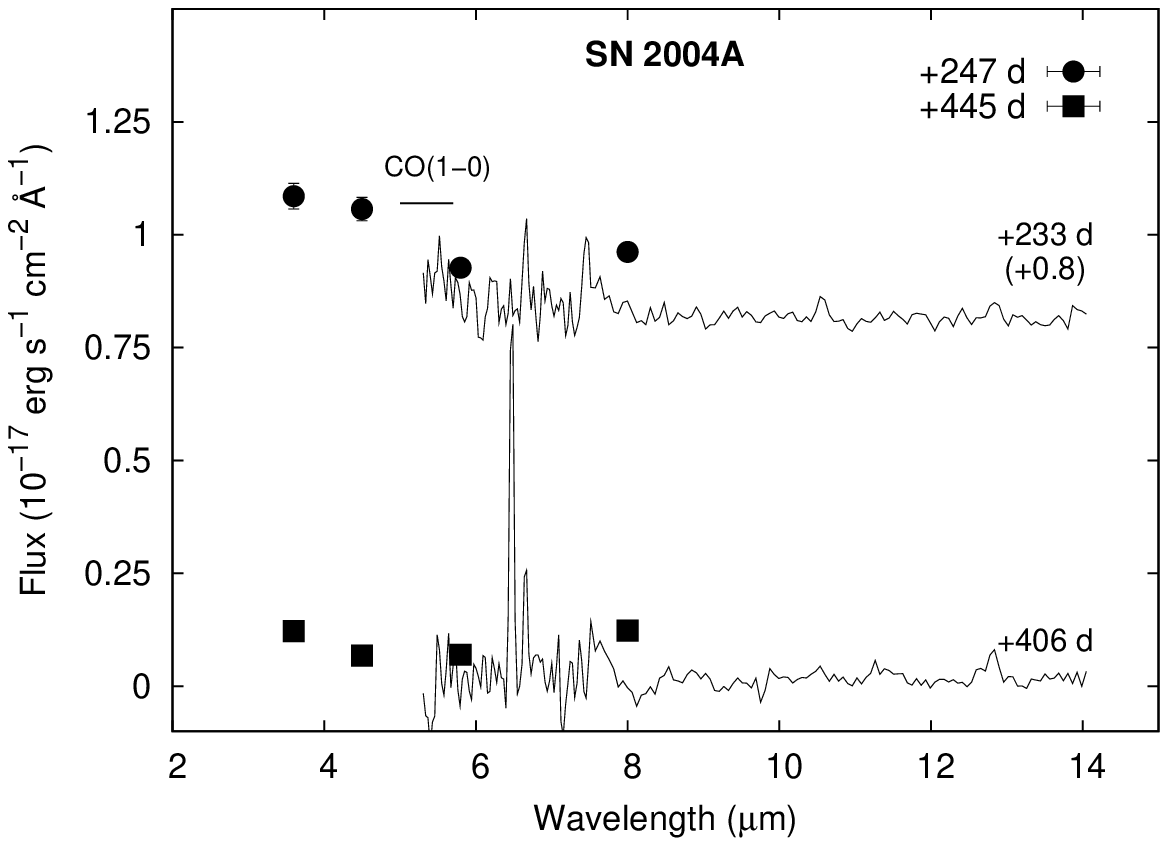}
\includegraphics[width=7.5cm, height=5.25cm]{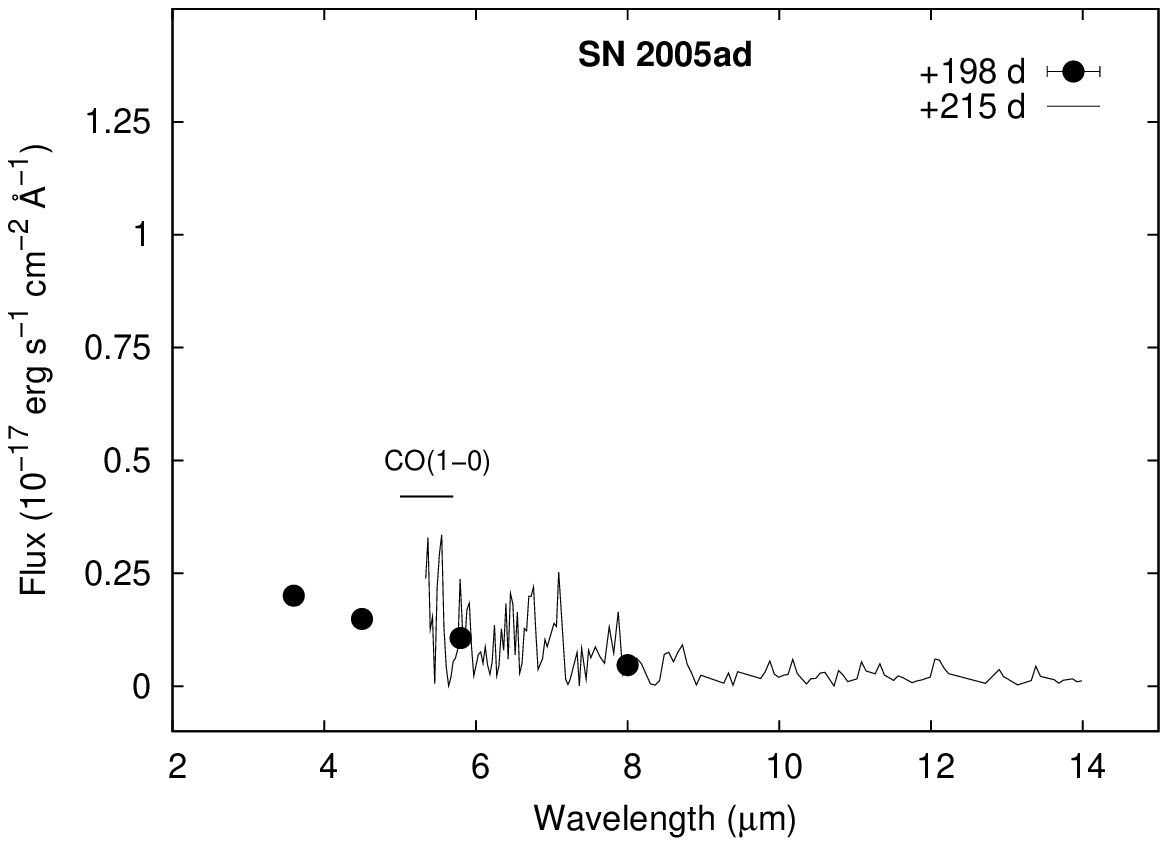}
\includegraphics[width=7.5cm, height=5.25cm]{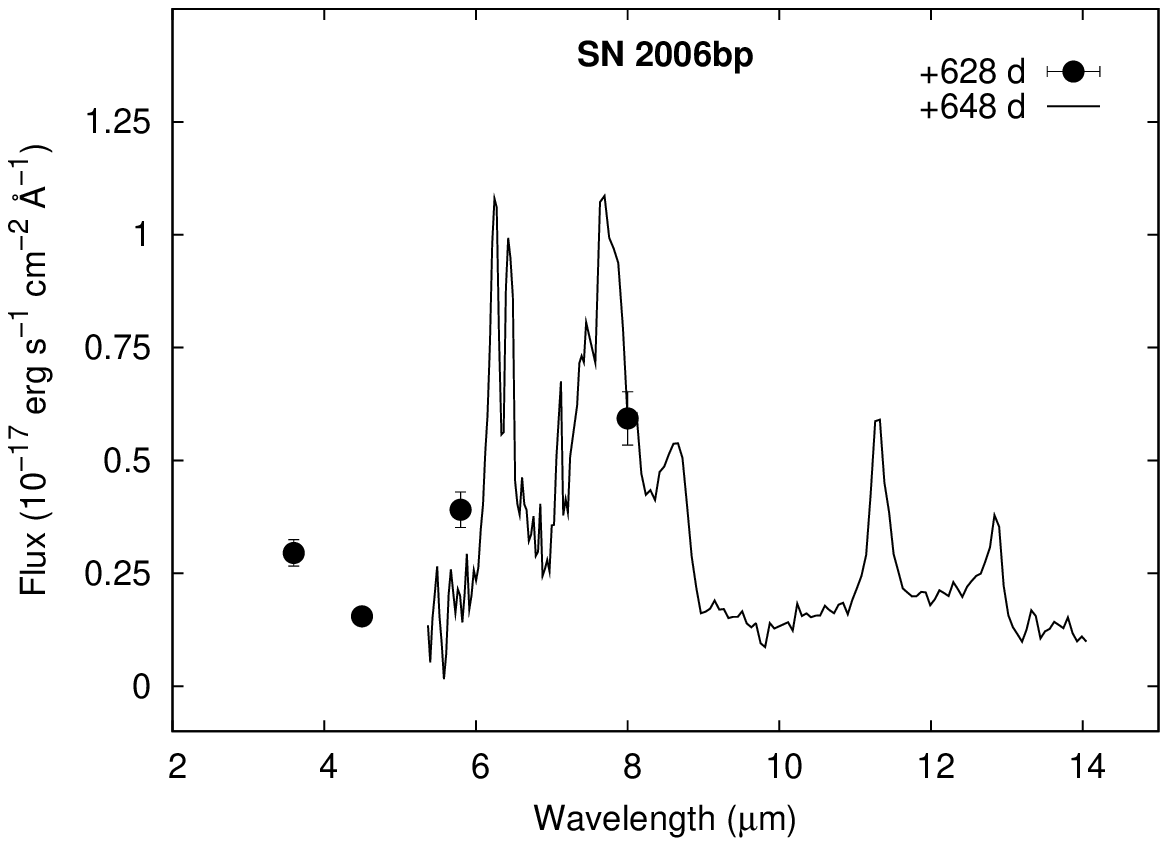}
\end{center}
\caption{IRS spectra of SNe 2003ie, 2004A, 2005ad, and 2006bp. The spectra of SN~2004A are vertically shifted relative to each other by the amount indicated 
in the label next to the upper spectrum. Strong and wide emission features in the spectrum of SN~2006bp possibly emerge from nearby interstellar clouds.}
\label{fig:irs}
\end{figure*}

We found that the spectra of SNe 2005cs and 2006ov are too noisy for further analysis. For the other SNe the spectral and photometric 
fluxes agree reasonably well. We also compared the extracted spectra with the public data in the Cornell Atlas of Spitzer IRS Sources
\footnote{The Cornell Atlas of Spitzer/IRS Sources (CASSIS) is a product of the Infrared Science Center at Cornell University, supported by NASA and JPL.} 
\citep[CASSIS,][]{Lebouteiller11}, and found them to be similar to each other.

The spectra of SN~2005af are shown in Figure \ref{fig:2005af_irs}. The three earliest spectra were already published by \citet{Kotak06} and \citet{Kotak08}. Our spectra 
agree well with theirs, which serves as a good test for checking our data reduction. In Figure \ref{fig:irs} we present the extracted IRS spectra for the other SNe.  
Although detailed modeling of these spectra is beyond the scope of this paper, we identified spectral features of SiO, and the 
1-0 vibrational transition of CO based on line identifications of \citet{Kotak05} and \citet{Kotak06}. The SiO feature plays an important role in modeling the 
observed SEDs, while the CO 1-0 transition results in a flux excess at 4.5 micron (see Section~\ref{anal}). 

\section{Analysis}\label{anal}

\subsection{Mid-IR SEDs}

\begin{figure*}
\centering
\includegraphics[width=6cm, height=4.14cm]{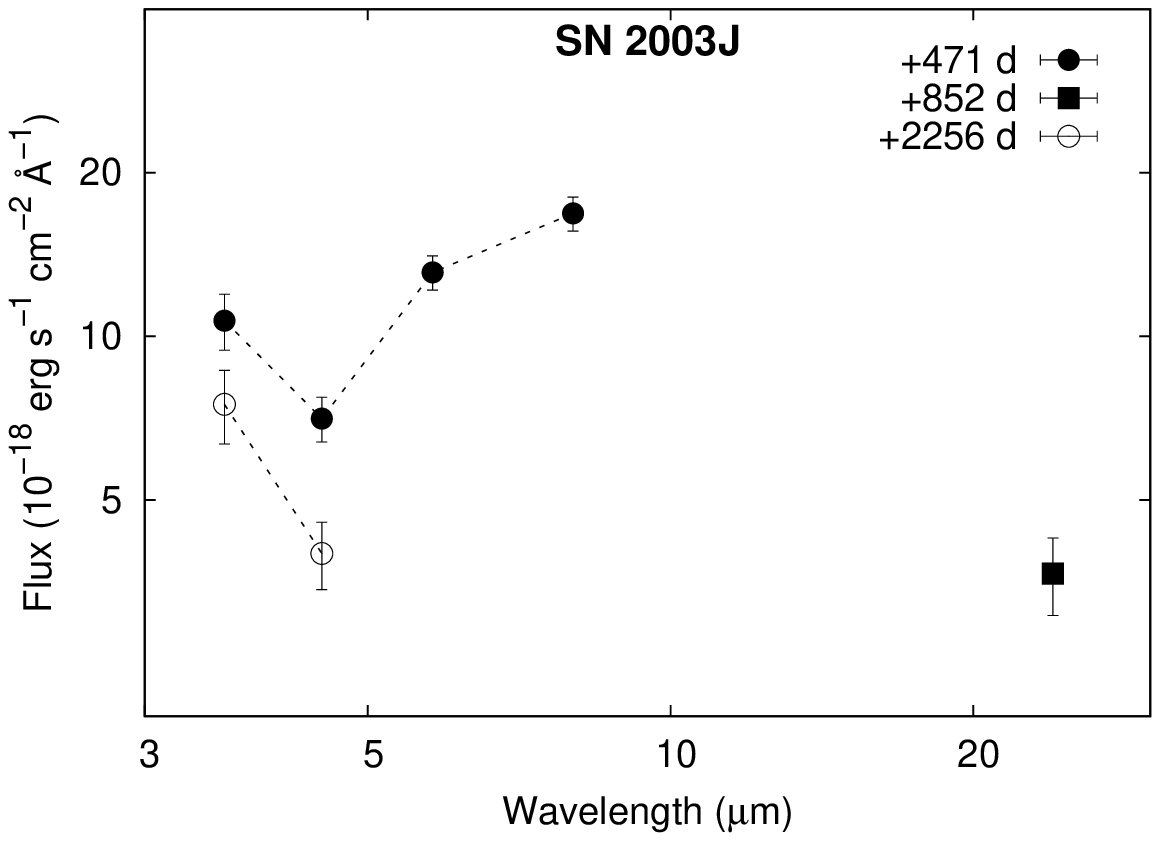} 
\includegraphics[width=6cm, height=4.14cm]{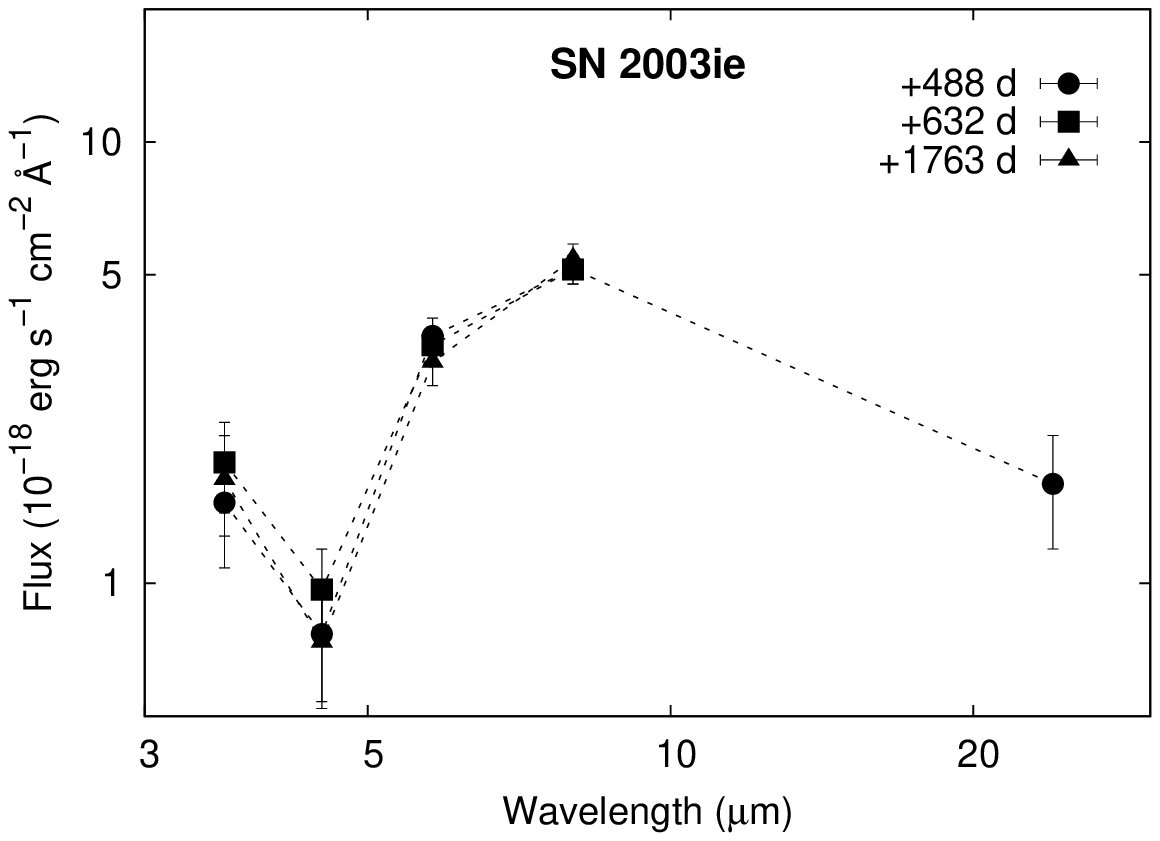} 
\includegraphics[width=6cm, height=4.14cm]{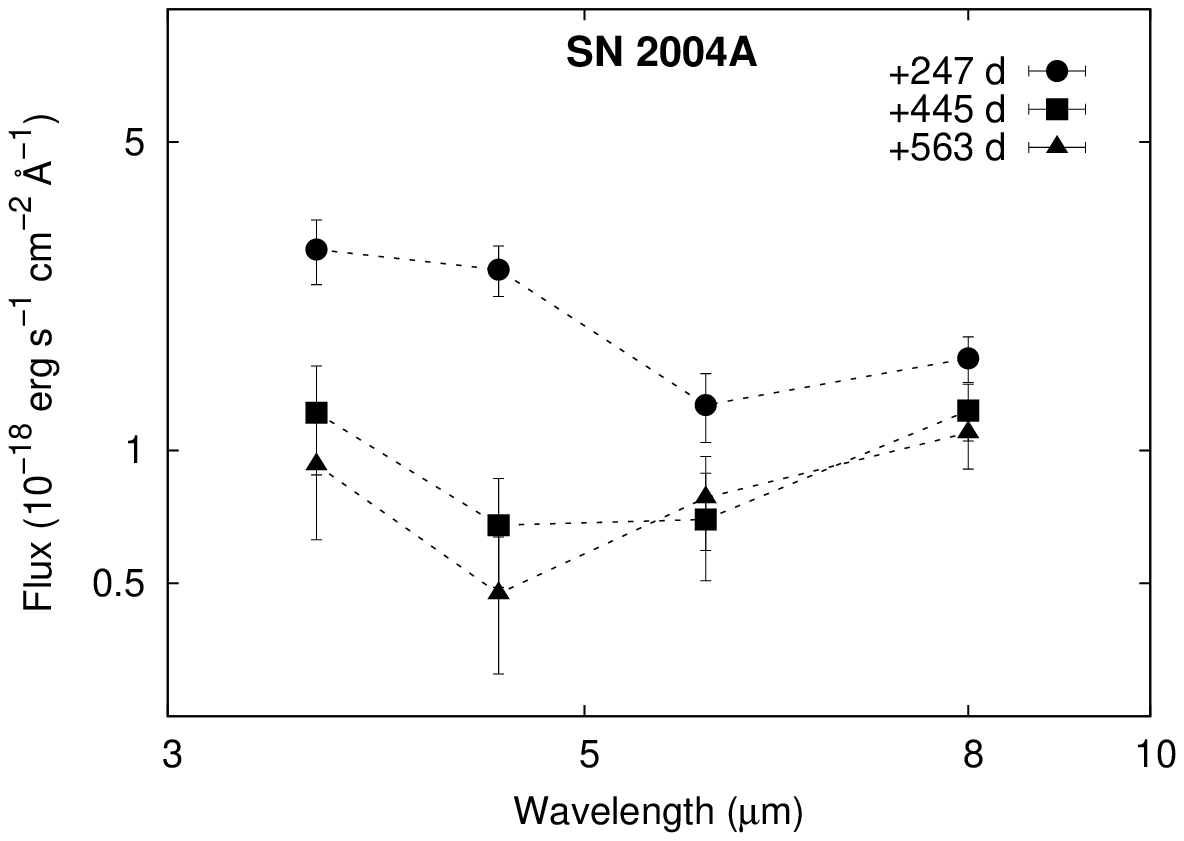} \vspace{3mm}
\includegraphics[width=6cm, height=4.14cm]{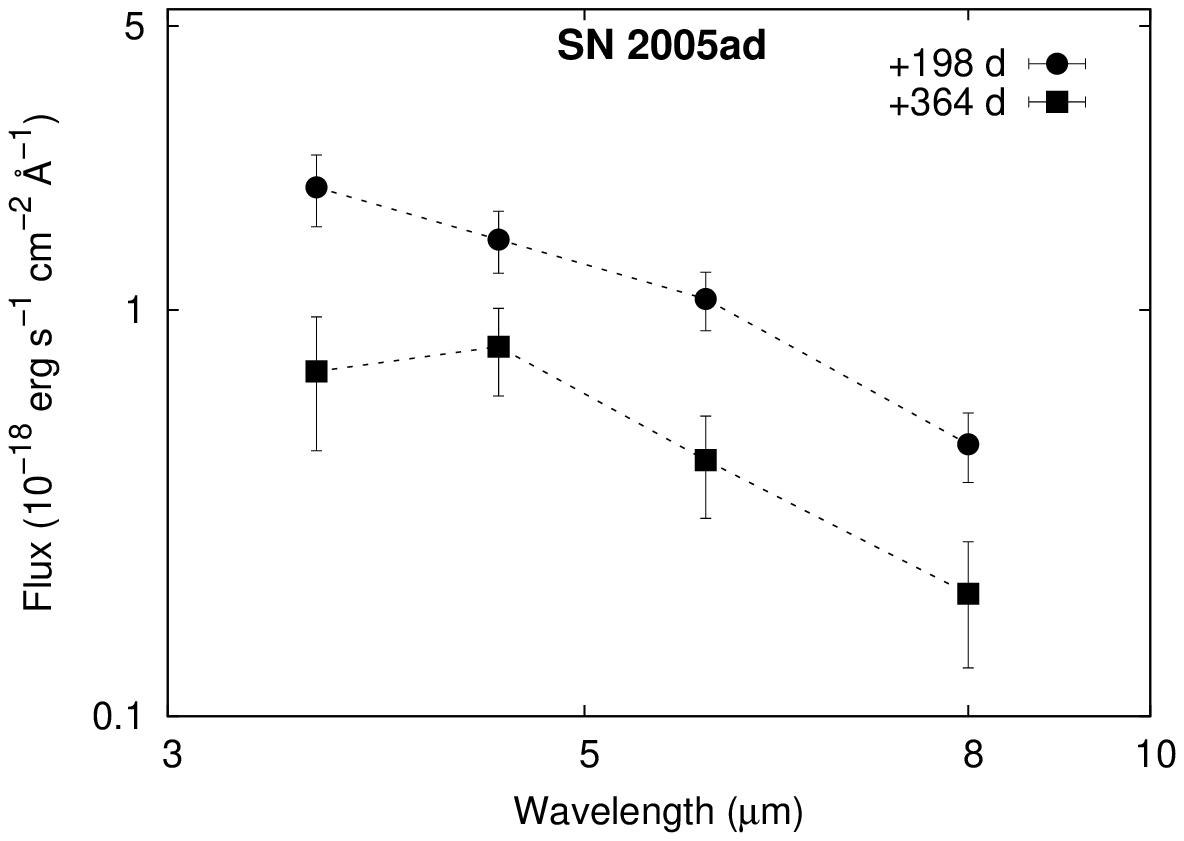}
\includegraphics[width=6cm, height=4.14cm]{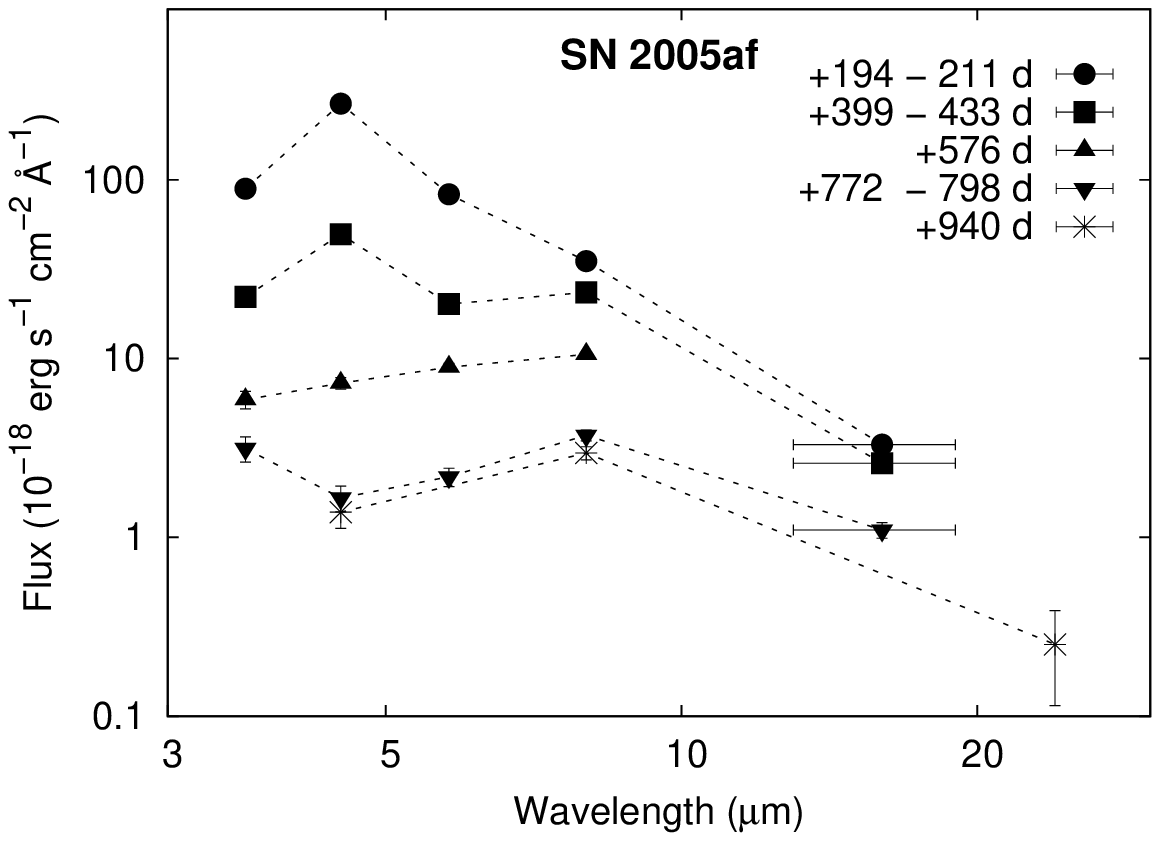}
\includegraphics[width=6cm, height=4.14cm]{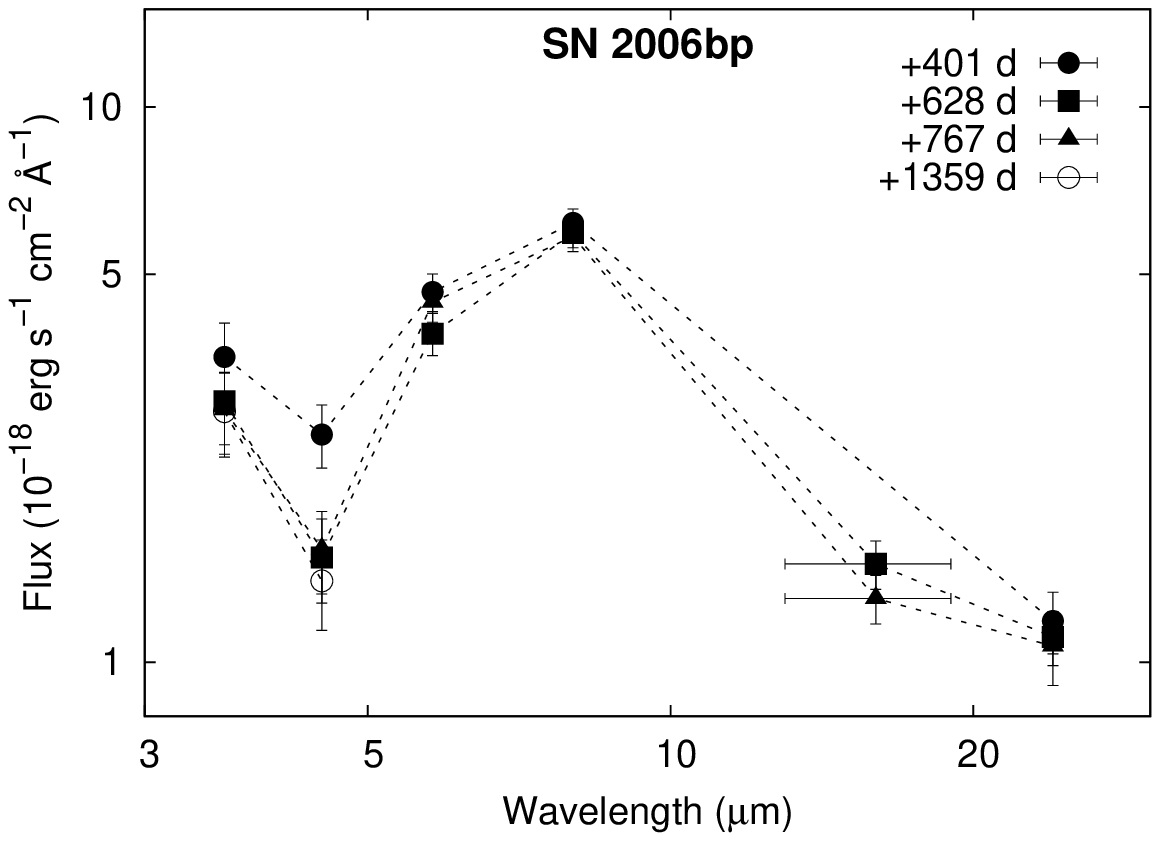} \vspace{3mm}
\includegraphics[width=6cm, height=4.14cm]{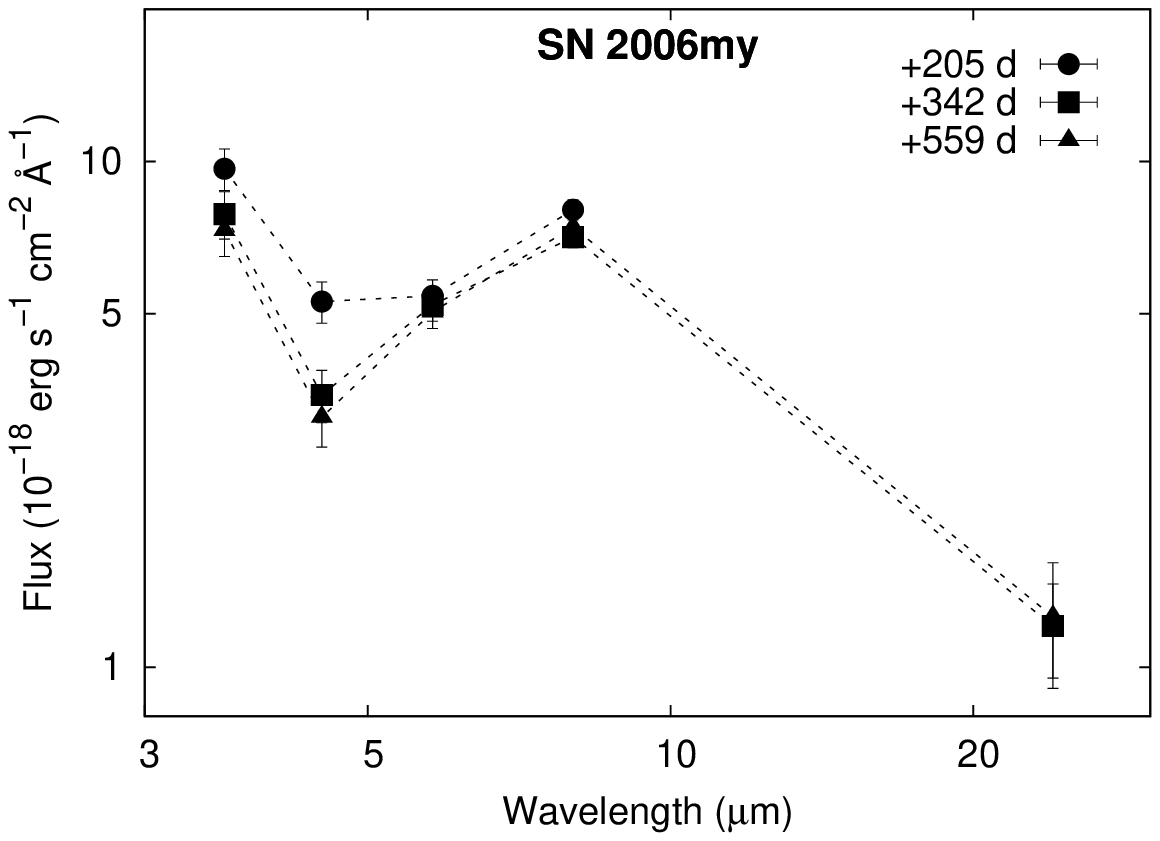}
\includegraphics[width=6cm, height=4.14cm]{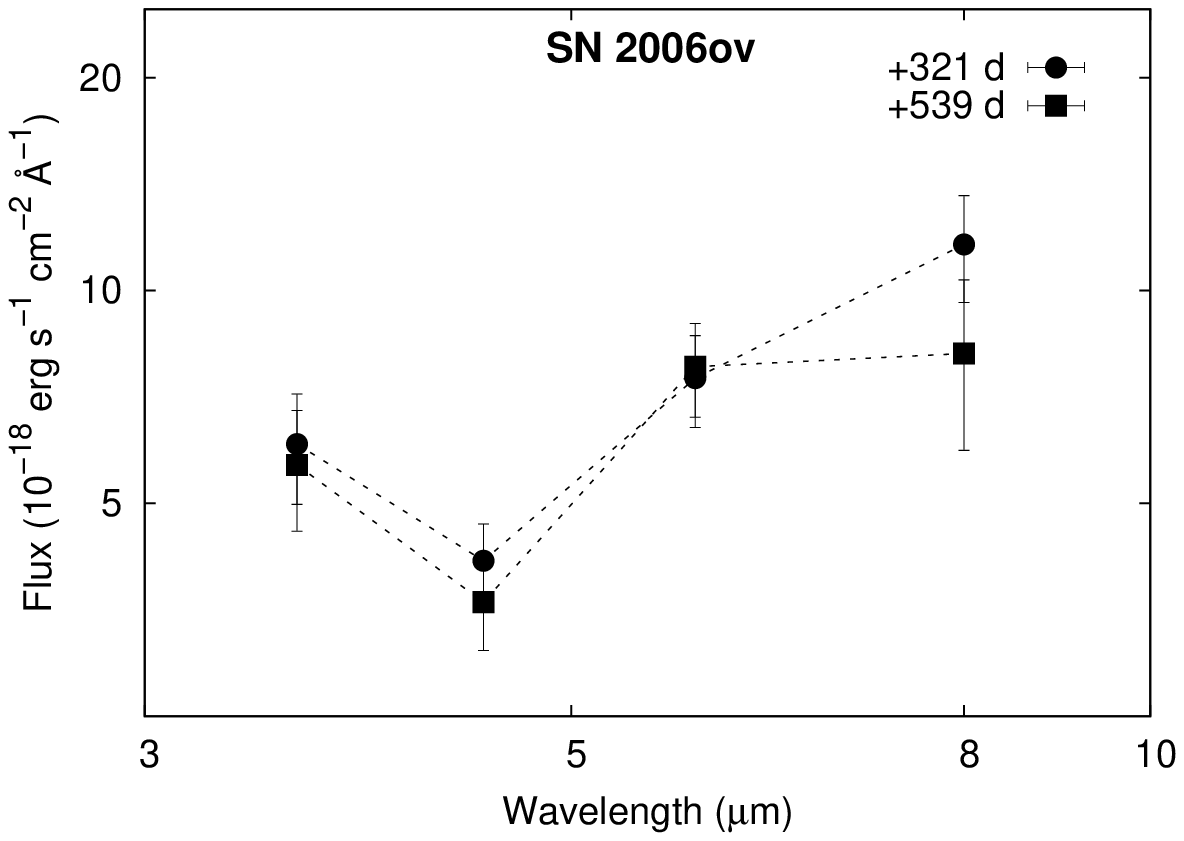}
\includegraphics[width=6cm, height=4.14cm]{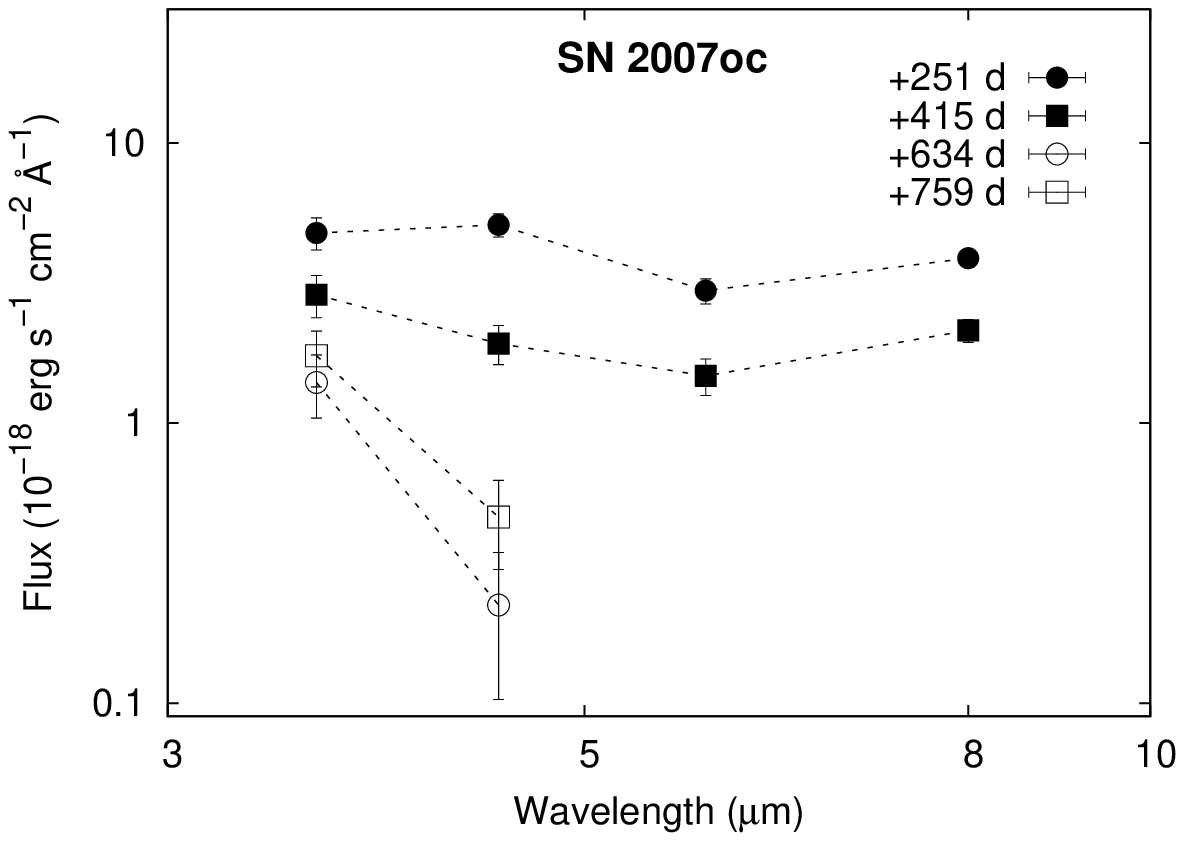}
\caption{Mid-IR SEDs of the nine CC SNe at different epochs. Open symbols denote IRAC warm mission data.}
\label{fig:sed}
\end{figure*}

Figure \ref{fig:sed} shows the mid-IR SEDs of the SNe calculated from IRAC 3.6, 4.5, 5.8, and 8.0 $\mu$m fluxes (plotted with available PUI 13.0-18.5 $\mu$m and MIPS 
24.0 $\mu$m fluxes). The continuum fluxes of the reduced IRS spectra match the SED flux levels well.

As is seen in Fig. \ref{fig:sed}, IRAC fluxes of several SNe (2003J, 2004A, 2005ad, 2005af, 2007oc) show obvious variations in time, while others do not. Based on the 
analysis of other CC SNe, e. g., SN~2004et \citep{Kotak09} or SN~2004dj \citep{Szalai11,Meikle11}, this temporal evolution could be the sign of newly-formed dust 
grains in the ejecta, but there may be more effects that have to be taken into account to explain this phenomenon.

\subsection{Fitting of dust models to MIR SEDs}\label{models}

Assuming that the radiation is purely thermal, the main source of mid-IR flux is most likely warm dust. To estimate the minimum
amount of dust, we fit analytic dust models to the observed SEDs. Prior to fitting, the observed SED fluxes were dereddened using the
galactic reddening law parametrized by \citet{Fitzpatrick07} assuming R$_V$ = 3.1 and adopting $E(B-V)$ values as listed in Section~\ref{sn}.

As a first step we examined whether simple, pure blackbody (BB) emission might be consistent with the observed MIR SEDs. 
Although a pure BB model has little physical significance, we applied it to estimate the {\it minimum} radius of an optically thick dust sphere that fits the observed 
SEDs. From the minimum radii we calculated the corresponding velocities assuming homologous expansion (see Table \ref{tab:bbpar}).
The typical velocities of type II-P SNe are 2-3000 km s$^{-1}$ for the inner ejecta where dust formation may take place. 
Therefore, for a larger minimum size it is not possible that the inner ejecta reach the calculated size of the BB. This 
means that in these cases the minimum radii of the optically thick dust spheres are too large and that consequently, most of the estimated dust mass cannot have originated from newly formed grains.

\begin{table}\small
\centering
\caption{\label{tab:bbpar} Minimum radii and corresponding velocities determined from the best-fit BB-models to observed SEDs.}
\newcommand\T{\rule{0pt}{3.1ex}}
\newcommand\B{\rule[-1.7ex]{0pt}{0pt}}
\begin{tabular}{llcc}
\hline
\hline
Object & Epoch & $R_{BB}$ & $v_{BB}$ \T \\
 & (days) & (10$^{16}$ cm) & km s$^{-1}$ \B \\
\hline
SN~2003J & 471$^{\dagger}$ & 6.3 & 15\,480\T \B\\
\hline
SN~2003ie & 488 & 6.0 & 14\,230\T \\
 & 632 & 4.9 & 8975 \\
 & 1763 & 6.1 & 4005\B \\
\hline
SN~2004A & 247 & 3.4 & 15\,930\T \\
 & 445 & 3.8 & 9885 \\
 & 563 & 2.2 & 4520\B \\
\hline
SN~2005ad & 198$^{\dagger}$ & 0.3 & 1755\T \\
 & 364$^{\dagger}$ & 0.2 & 635\B \\
\hline
SN~2005af & 194 & 0.8 & 4770\T \\
 & 399 & 0.4 & 1160 \\
 & 576 & 0.6 & 1205 \\
 & 772 & 0.8 & 1200 \\
 & 940 & 0.6 & 740\B \\
\hline
SN~2006bp & 401 & 4.5 & 12\,990\T \\
 & 628 & 5.5 & 10\,140 \\
 & 767 & 5.1 & 7695\B \\
\hline
SN~2006my & 205 & 3.8 & 21\,455\T \\
 & 342 & 3.7 & 12\,520 \\
 & 559 & 4.2 & 8695\B \\
\hline
SN~2006ov & 321 & 5.1 & 18\,390\T \\
 & 539 & 3.4 & 7300\B \\
\hline
SN~2007oc & 250$^{\dagger}$ & 5.7 & 26\,285\T \\
 & 415$^{\dagger}$ & 4.9 & 13\,665\B \\
\hline
\end{tabular}
\tablefoot{
$^{\dagger}$ From date of discovery.}
\end{table}

The analytic models were calculated using Eq.1 in \citet{Meikle07} assuming a homogeneous (constant-density) dust distribution:

\begin{equation}
L_{\nu}=2 \pi^2 R^2 B_{\nu}(T)\left[ \tau_{\nu}^{-2} (2\tau_{\nu}^2 -1 + (2 \tau_{\nu} + 1)\exp(-2 \tau_{\nu}) \right],
\end{equation}

\noindent where $R$ is the radius of the dust sphere, $B_{\nu}(T)$ is the Planck function, $T$ is the temperature of the dust, and $\tau_{\nu}$ is the optical depth at a 
given frequency. 
To estimate the dust optical depth, we adopted the power-law grain-size distribution of \citet{Mathis77}, hereafter MRN, assuming $m$ = 3.5 for the power-law index and
grain sizes between $a_{min}$ = 0.005 $\mu$m and $a_{max}$ = 0.05 $\mu$m. $\tau_{\nu}$ can then be calculated as

\begin{equation}
\tau_{\nu} = \frac{4}{3} \pi k \rho_{grain} \kappa_{\nu} R \frac{1}{4-m} \left[ a^{4-m}_{max} - a^{4-m}_{min} \right],
\end{equation}

\noindent where $\rho_{grain}$ is the density of dust grain material, $k$ is the grain number-density scaling factor, and $\kappa_{\nu}$ is the mass absorption coefficient.
The dust is assumed to be distributed uniformly within a sphere. $R$, $T$, and $k$ were free parameters during the fitting.

\begin{figure*}
\centering
\includegraphics[width=6cm, height=4.14cm]{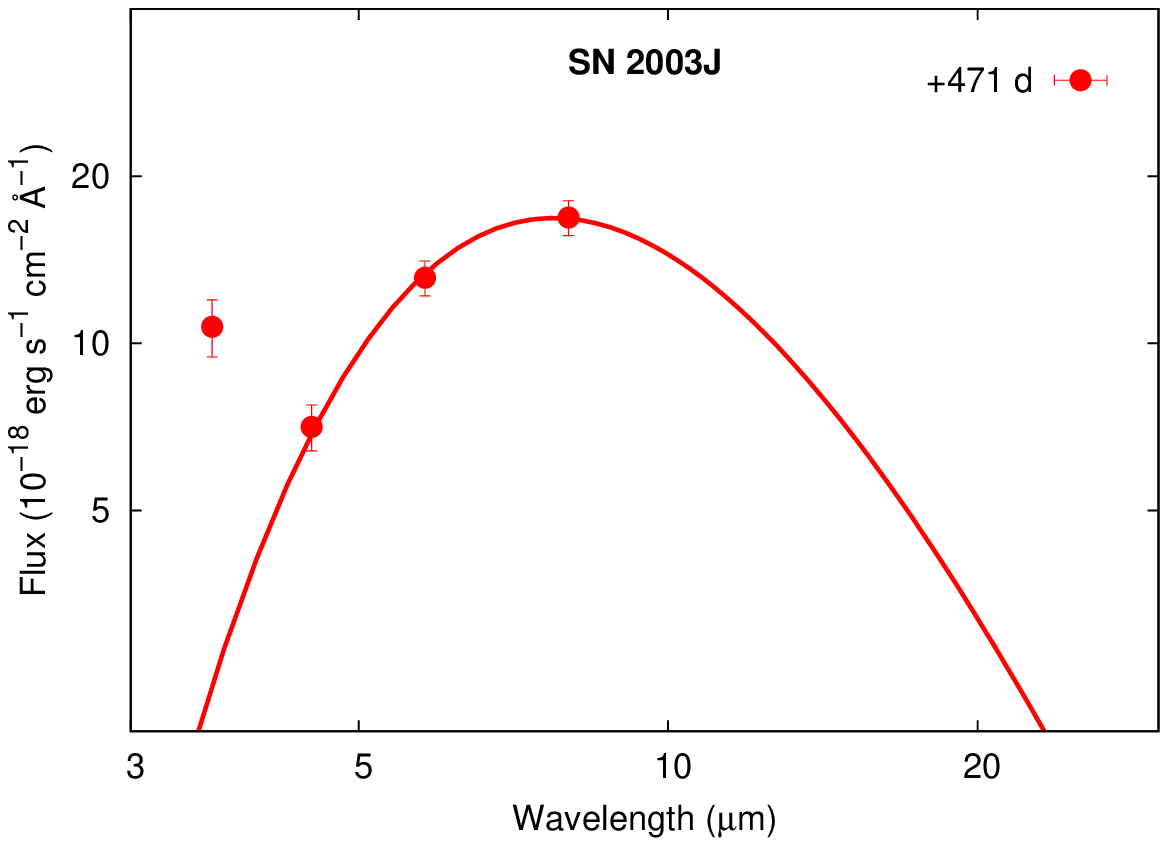} 
\includegraphics[width=6cm, height=4.14cm]{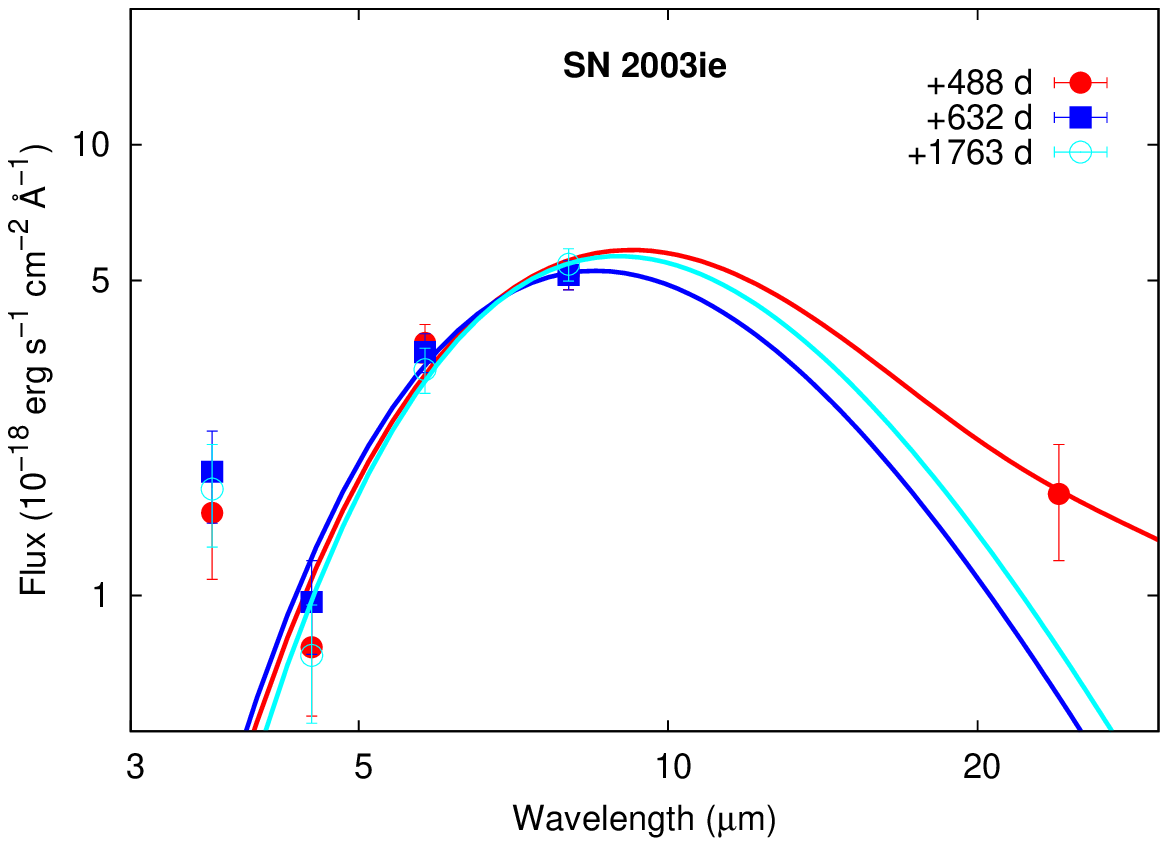} 
\includegraphics[width=6cm, height=4.14cm]{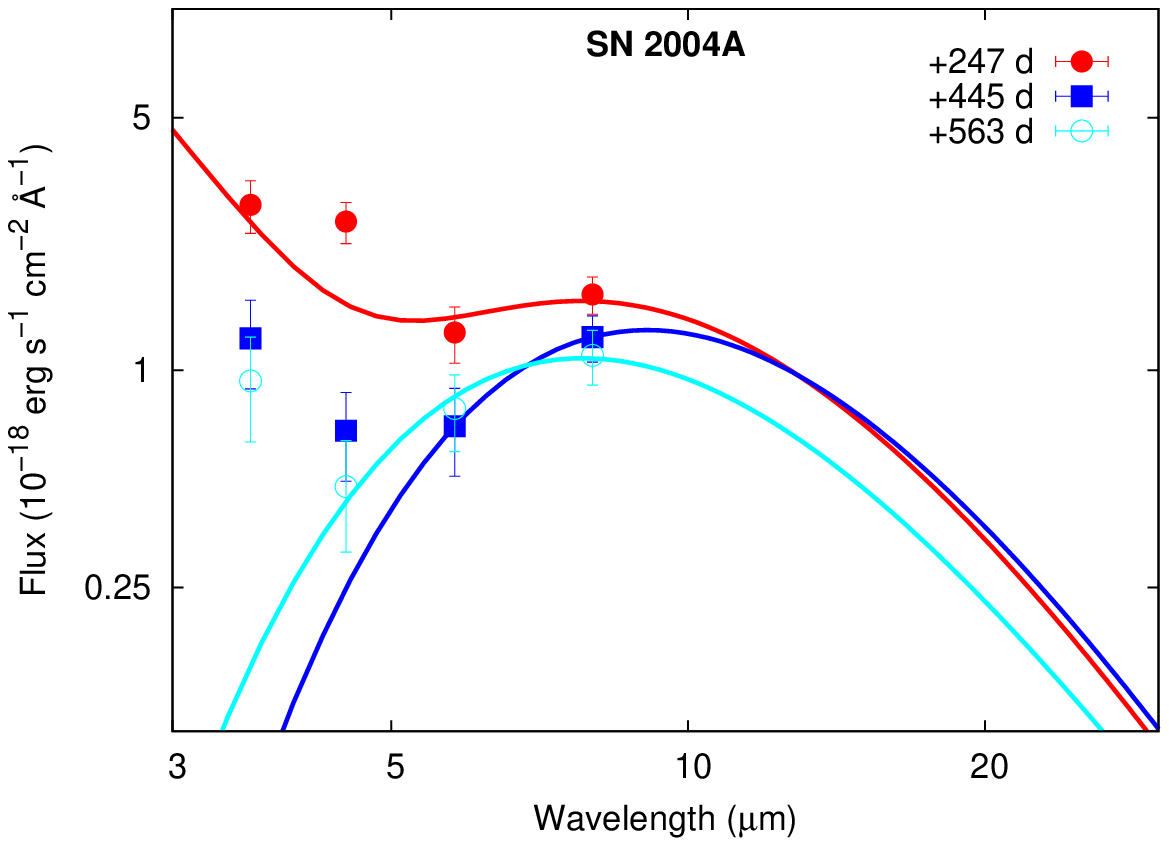} \vspace{3mm}
\includegraphics[width=6cm, height=4.14cm]{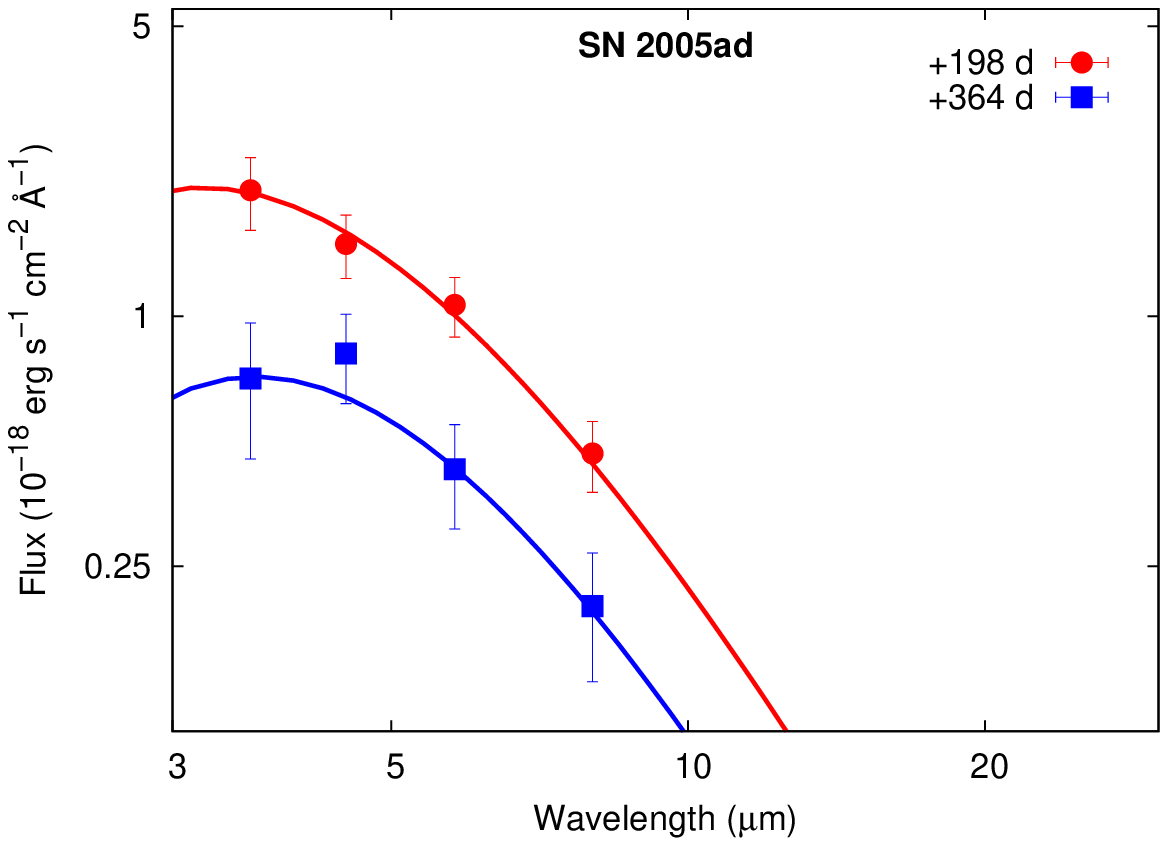} 
\includegraphics[width=6cm, height=4.14cm]{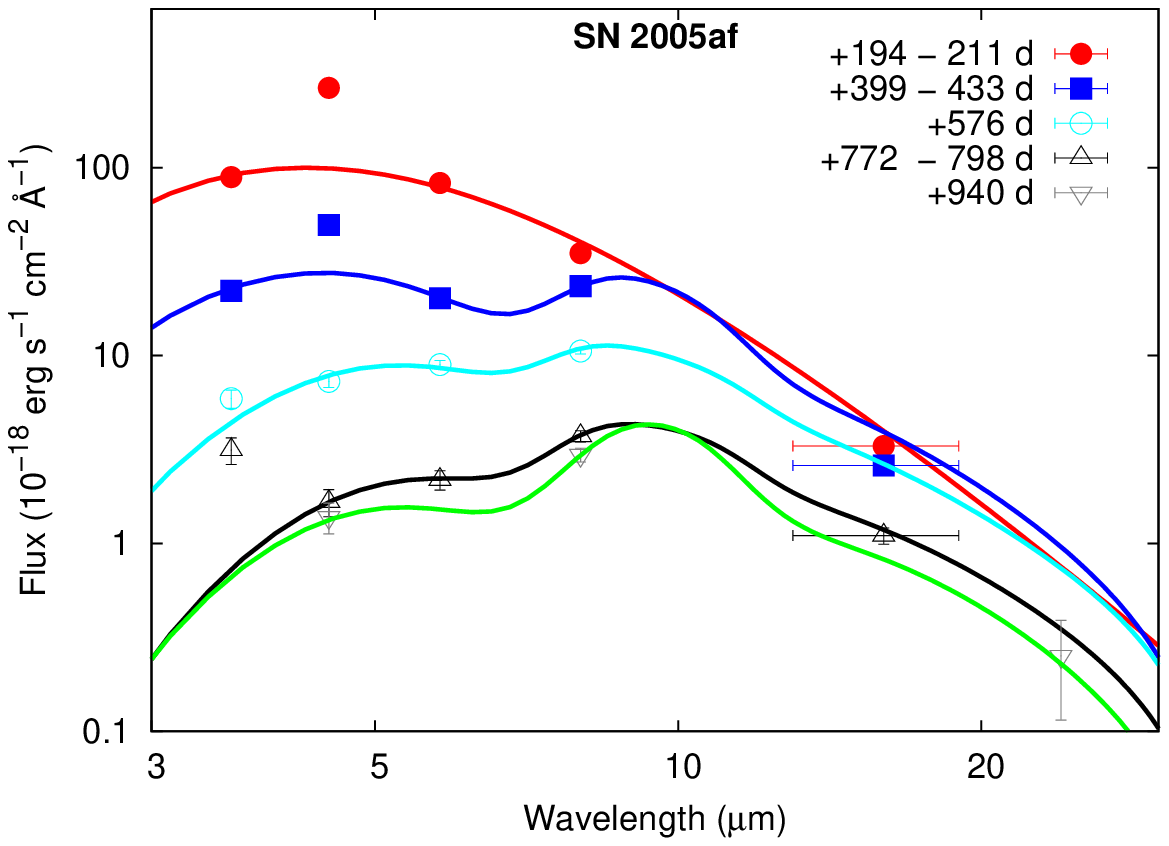}
\includegraphics[width=6cm, height=4.14cm]{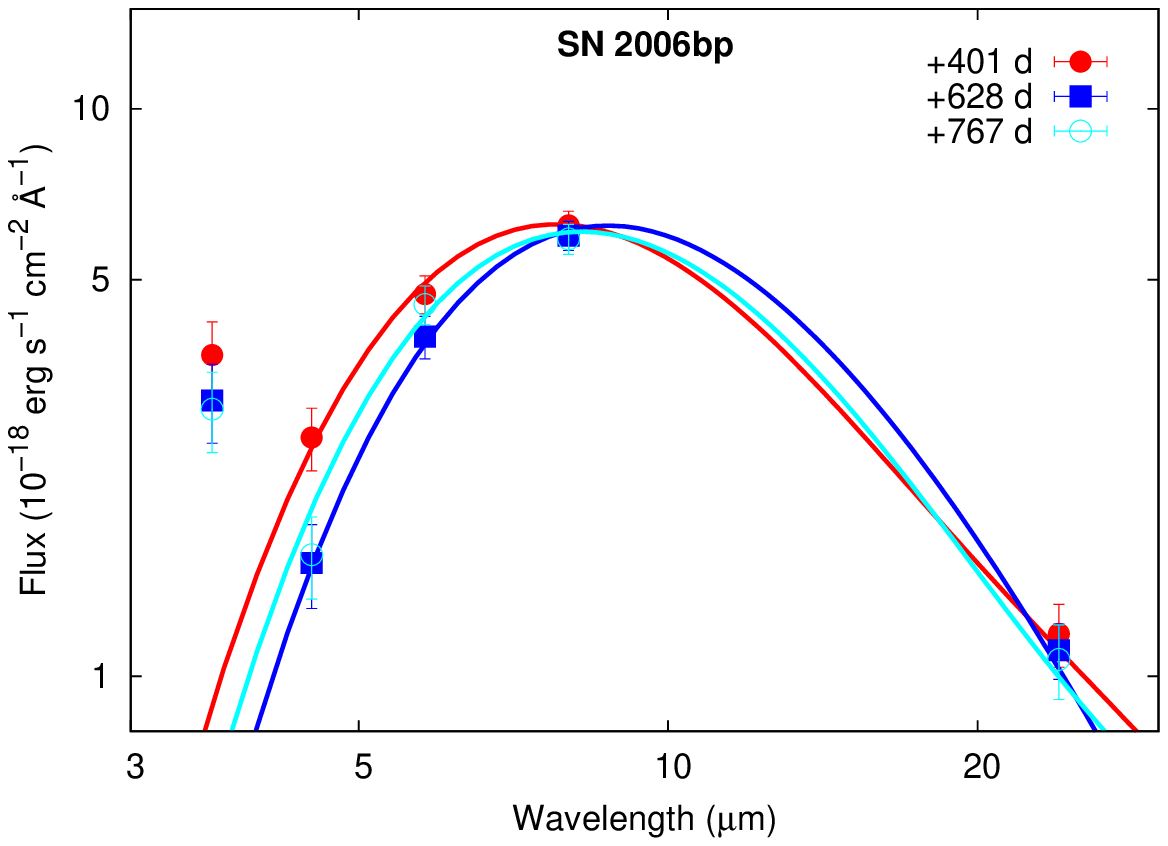} \vspace{3mm}
\includegraphics[width=6cm, height=4.14cm]{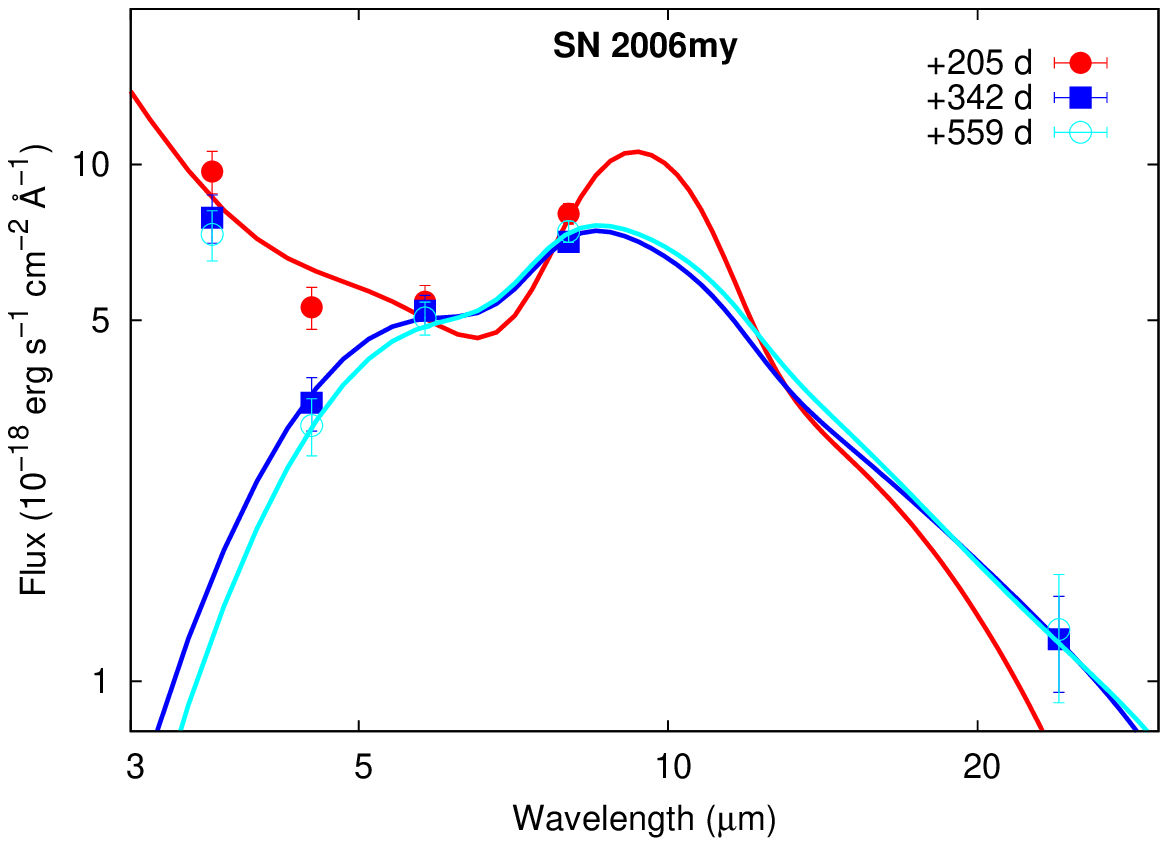}
\includegraphics[width=6cm, height=4.14cm]{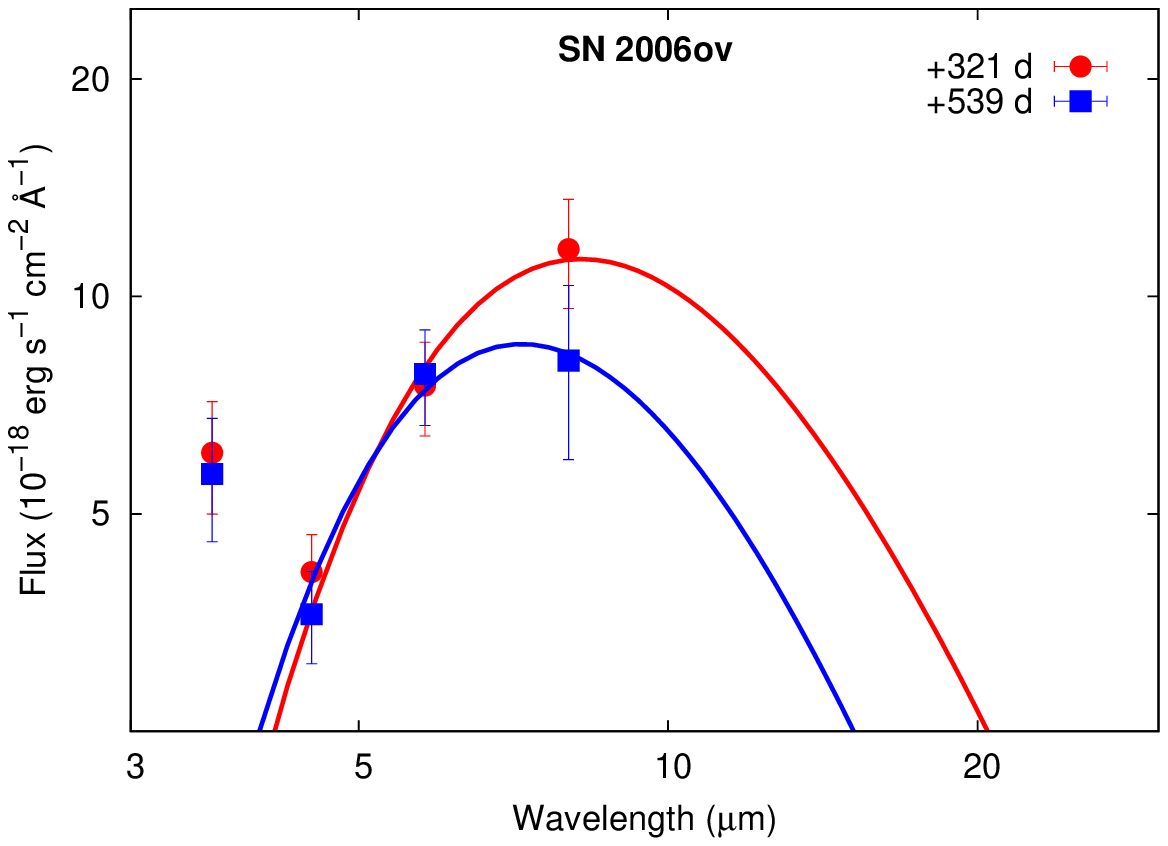}
\includegraphics[width=6cm, height=4.14cm]{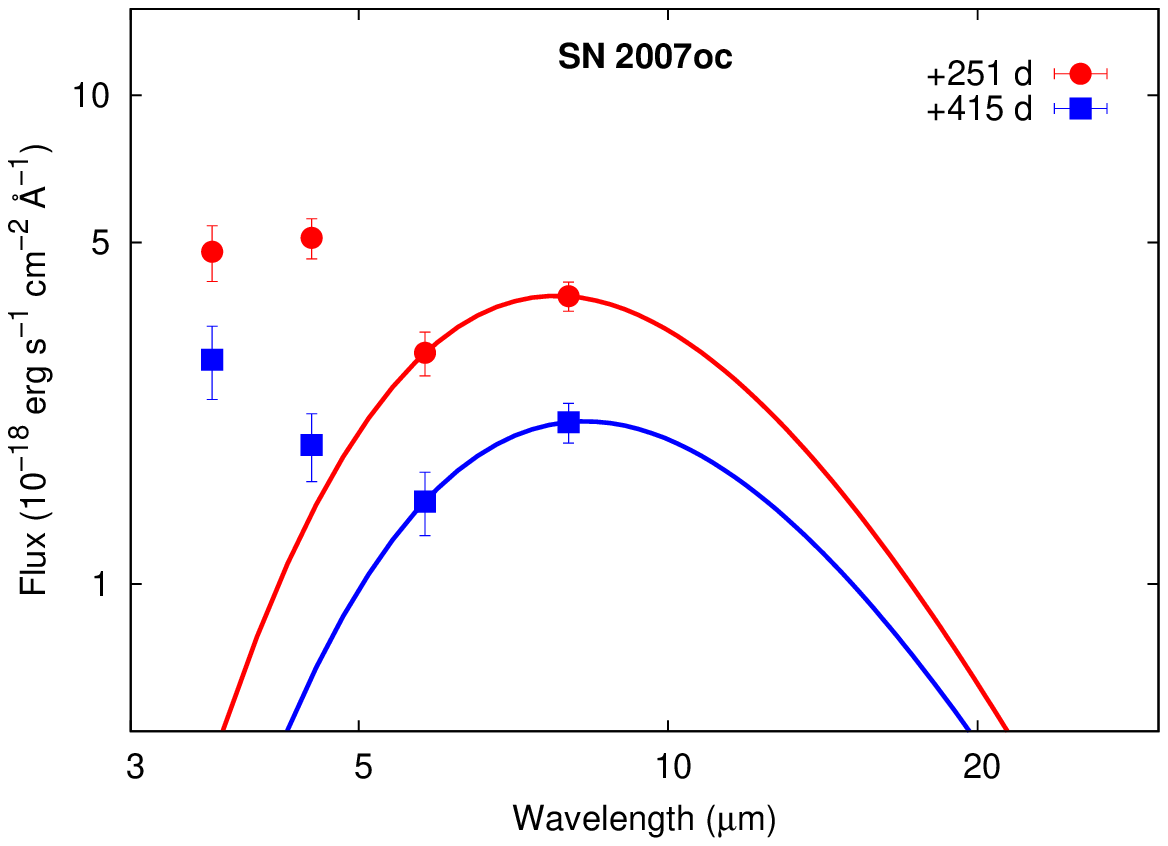}
\caption{Best-fit analytic models on MIR SEDs of the SNe. The 3.6 $\mu$m and sometimes the 4.5 $\mu$m fluxes are usually not included in the fitting (see text). 
The fittings that contain hot and warm components for SN~2004A and SN~2006my are shown in Figure \ref{fig:hot}.} 
\label{fig:models}
\end{figure*}

First we calculated the dust models using amorphous carbon (AC) grains. Dust opacity values were taken from \citet{Colangeli95}, while the density of the dust grain material, 
$\rho_{grain}$ = 1.85 g cm$^{-3}$, was adopted from \citet{Rouleau91}. Although we were able to successfully apply these models in most cases, finding adequate solutions for SNe 
2005af and 2006my was possible only by assuming silicate dust \citep[C-Si-PAH mixture, MRN-distribution, $\rho_{grain}$ = 1.85 g cm$^{-3}$,][]{Weingartner01}. The IRS 
spectra of SN~2005af also show the signs of SiO \citep[see Section~\ref{obs} and][]{Kotak06}; but there were no IRS spectra on SN~2006my in the database.
On the other hand, the spectra of the other SNe (2003ie, 2004A, 2005ad, 2006bp) that can be fit well with AC models do not show obvious signs of SiO.
The parameters of the best-fit analytic models (shown in Figure \ref{fig:models}) are collected in Tables \ref{tab:dustpar} and \ref{tab:dustpar2}.

\begin{table}\small
\begin{center}
\caption{\label{tab:dustpar} Parameters for the best-fit analytic amorphous carbon dust models to the warm components of SEDs.}
\newcommand\T{\rule{0pt}{3.1ex}}
\newcommand\B{\rule[-1.7ex]{0pt}{0pt}}
\begin{tabular}{cccccc}
\hline
\hline
Epoch & $T_{warm}$ & $R_{warm}$ & $k$ & $M_{dust}$ & $L_{warm}$\T \\
(days) & (K) & (10$^{16}$ cm) & (10$^{-22}$) & (10$^{-5}$ M$_{\odot}$) & ($10^{39}$ erg s$^{-1}$)\B \\
\hline
\multicolumn{6}{c}{SN~2003J}\T \\	
\hline
471$^{\dagger}$ & 370 & 6.4 & 54.5 & 710 & 48.8\T \\
\hline
\multicolumn{6}{c}{SN~2003ie}\T \\
\hline
488$^{\ddagger}$ & 310 & 6.2 & 60 & 700 & 22.2\T \\
632 & 300 & 10.1 & 4 & 200 & 16.9 \\
1763 & 280 & 15.8 & 1.5 & 290 & 19.1 \\
\hline
\multicolumn{6}{c}{SN~2004A}\T \\
\hline
247$^{\ddagger}$ & 340 & 3.4 & 100 & 160 & 8.5\T \\
445 & 310 & 3.9 & 67 & 200 & 8.1 \\
563 & 370 & 2.2 & 350 & 180 & 6.2 \\
\hline
\multicolumn{6}{c}{SN~2005ad}\T \\
\hline
198$^{\dagger}$ & 890 & 0.4 & 480 & 1.0 & 4.9\T \\
364$^{\dagger}$ & 750 & 0.4 & 190 & 0.4 & 1.8 \\
\hline
\multicolumn{6}{c}{SN~2005af}\T \\
\hline
194 & 590 & 2.3 & 7.0 & 4.2 & 10.2 \T \\
399 & 640 & 0.6 & 190 & 1.6 & 3.3 \\
576 & 460 & 0.9 & 85 & 2.6 & 1.5 \\
772 & 430 & 0.8 & 66 & 0.7 & 0.3 \\
940 & 380 & 0.7 & 260 & 4.5 & 0.6 \\
\hline
\multicolumn{6}{c}{SN~2006bp}\T \\
\hline
401$^{\ddagger}$ & 370 & 4.6 & 100 & 480 & 26.2\T \\
628 & 330 & 6.1 & 89 & 1000 & 29.3 \\
767$^{\ddagger}$ & 350 & 5.2 & 100 & 690 & 27.0 \\
\hline
\multicolumn{6}{c}{SN~2006my}\T \\
\hline
205$^{\ddagger}$ & 470 & 3.8 & 110 & 88 & 19.3\T \\
342$^{\ddagger}$ & 420 & 4.1 & 67 & 230 & 32.5 \\
559$^{\ddagger}$ & 410 & 4.3 & 93 & 370 & 34.3 \\
\hline
\multicolumn{6}{c}{SN~2006ov}\T \\
\hline
321 & 350 & 5.1 & 100 & 650 & 25.9\T \\
539 & 350 & 8.2 & 3.0 & 82 & 15.4 \\
\hline
\multicolumn{6}{c}{SN~2007oc}\T \\
\hline
250$^{\dagger}$ & 340 & 8.9 & 9.0 & 310 & 37.8\T \\
415$^{\dagger}$ & 340 & 5.5 & 45 & 370 & 23.5 \\
\hline
\end{tabular}
\tablefoot{
$^{\dagger}$ From date of discovery. \\
$^{\ddagger}$ An additional (cold or hot) component was also fit (see details in Table \ref{tab:hotpar} and in the text).}
\end{center}
\end{table}

\begin{table}\small
\begin{center}
\caption{\label{tab:dustpar2} Parameters for the best-fit analytic silicate dust models to the warm components of SEDs.}
\newcommand\T{\rule{0pt}{3.1ex}}
\newcommand\B{\rule[-1.7ex]{0pt}{0pt}}
\begin{tabular}{cccccc}
\hline
\hline
Epoch & $T_{warm}$ & $R_{warm}$ & $k$ & $M_{dust}$ & $L_{warm}$\T \\
(days) & (K) & (10$^{16}$ cm) & (10$^{-22}$) & (10$^{-5}$ M$_{\odot}$) & ($10^{39}$ erg s$^{-1}$)\B \\
\hline
\multicolumn{6}{c}{SN~2005af}\T \\
\hline
399 & 550 & 1.1 & 200 & 21 & 4.3\T \\
576 & 450 & 1.0 & 450 & 31 & 2.0 \\
772 & 390 & 0.8 & 410 & 17 & 0.7 \\
940 & 400 & 0.8 & 160 & 8.1 & 0.5 \\
\hline
\multicolumn{6}{c}{SN~2006my}\T \\
\hline
205$^{\dagger}$ & 380 & 8.0 & 22 & 920 & 44.9\T \\
342$^{\dagger}$ & 400 & 5.5 & 100 & 1300 & 33.8 \\
559$^{\dagger}$ & 380 & 6.2 & 100 & 1900 & 28.4\B \\
\hline
\end{tabular}
\tablefoot{
$^{\dagger}$ An additional (cold or hot) component was also fit (see details in Table \ref{tab:hotpar} and in the text).}
\end{center}
\end{table}

\begin{figure*}
\centering
\includegraphics[width=7.5cm, height=5.25cm]{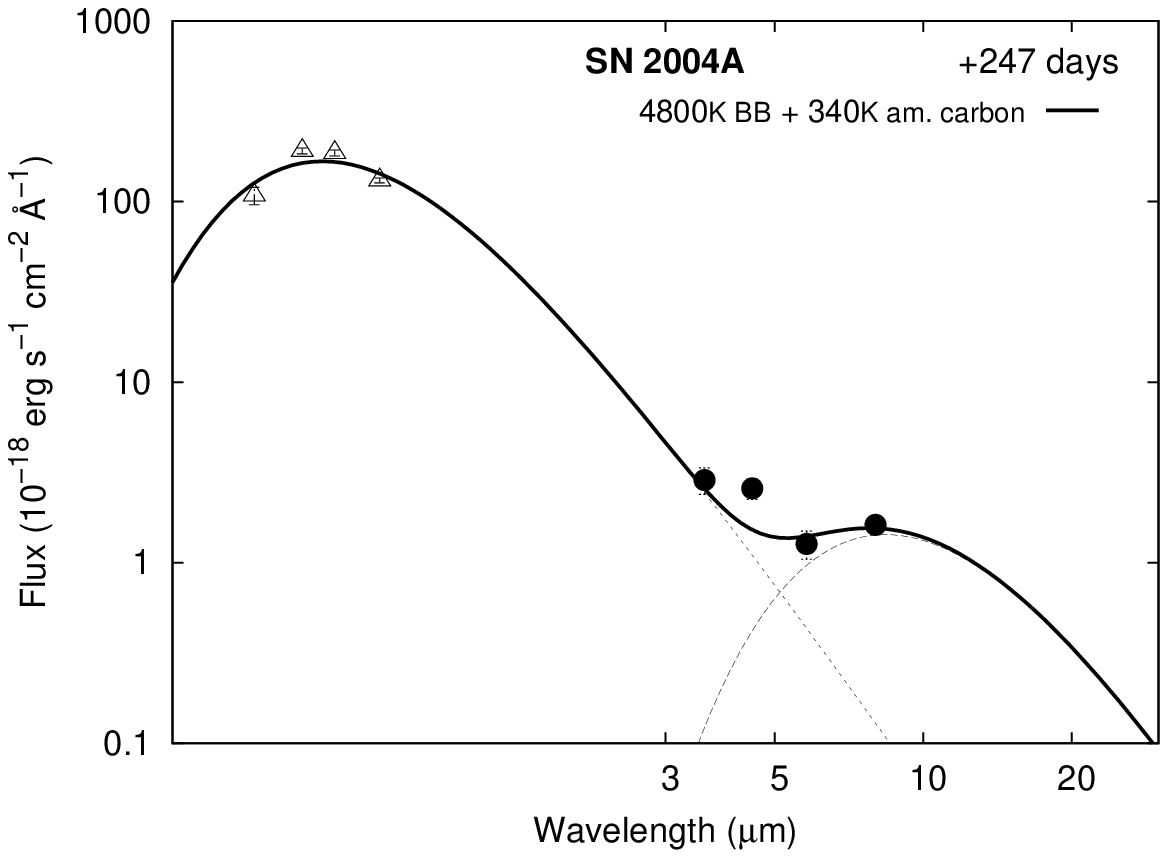} 
\includegraphics[width=7.5cm, height=5.25cm]{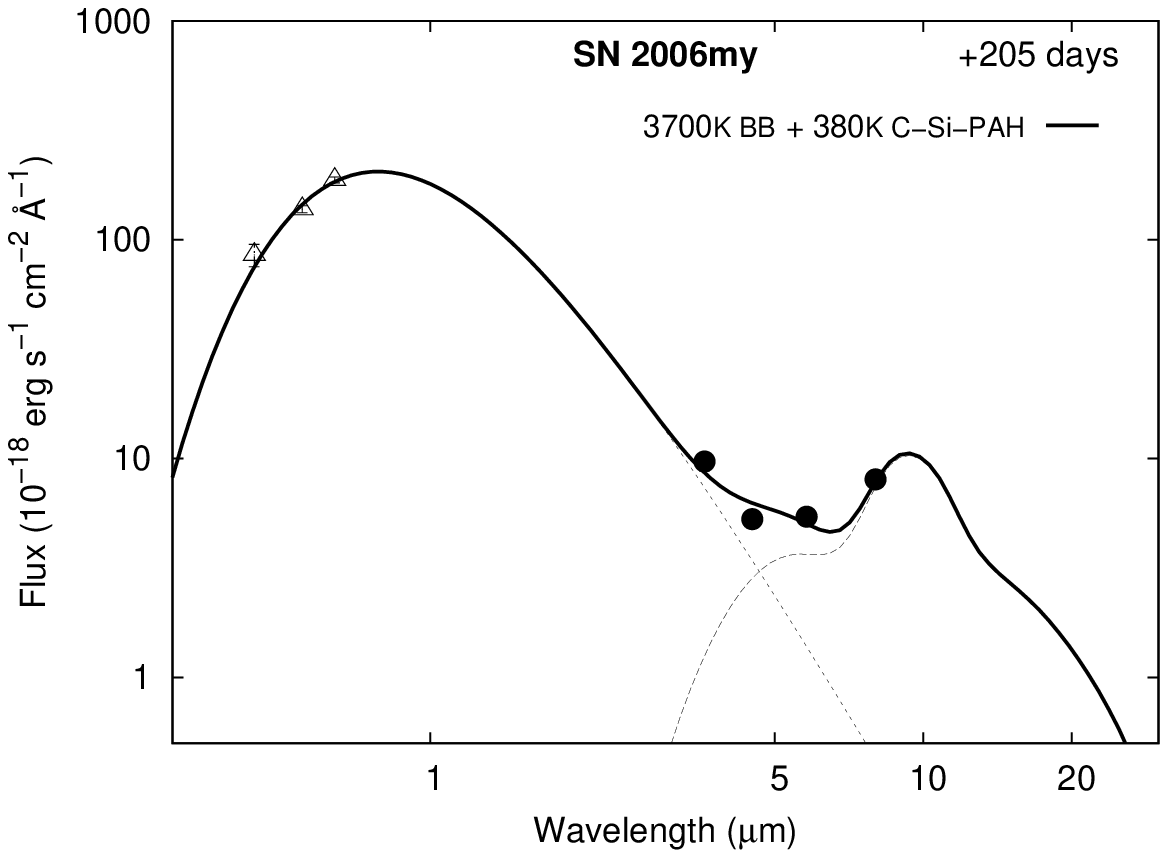} 
\caption{Best-fit models for SN~2004A at 247 days (left) and for SN~2006my at 205 days (right) that contain a warm dust component and a hot blackbody (see text).}
\label{fig:hot}
\end{figure*}

In a few cases we found that a single temperature model cannot describe the whole observed SED. This is similar to the result of \citet{Wooden93}, who showed that for 
SN~1987A a hot component ($T \sim$ 5-10\,000 K) may affect the continuum emission 
of the warm dust. This hot component is thought to be caused by an optically thick gas in the innermost part of the ejecta. 
The contribution of such a hot component can be significant at shorter IRAC wavelengths. The modeling of the hot component needs late-time optical or 
near-IR data taken simultaneously with the {\it Spitzer} observations.
Because of the lack of such data for most of the SNe in our sample, we usually did not include the 3.6 $\mu$m (and sometimes also 4.5 $\mu$m) fluxes while fitting 
the dust models. 
There are two exceptions for which we have found optical 
photometry obtained close to the IRAC measurements: SN~2004A at 247 days \citep{Hendry06}, and SN~2006my at 205 days \citep{Maguire10}. At these epochs we completed the 
observed SEDs with the optical data and added a hot component to our models (see the fittings in Figure \ref{fig:hot}, and the parameters in Table \ref{tab:hotpar}.).

Figure \ref{fig:models} shows that SNe 2004A, 2005ad, 2005af, and 2007oc show variable excess flux at 4.5 $\mu$m (in these cases we omitted 
the 4.5 $\mu$m fluxes from the fitting).
The excess flux weakens with age, as the
ejecta temperature decreases. This trend is also seen in the IRS spectra of SN~2005af \citep[see Section~\ref{obs} and][]{Kotak06}.
This may be due to the 
influence of the red wing of the 1-0 vibrational transition of CO at 4.65 $\mu$m \citep[see also][]{Kotak05,Kotak06,Szalai11}. An alternative explanation for temporarily 
changing flux at 4.5 $\mu$m may be the rapid destruction of stochastically heated grains. This effect becomes significant when the grain sizes are below 0.003 $\mu$m 
\citep{Bocchio12}. However, as we mentioned above, in the cases studied here the grain sizes are thought to be more than 0.005 $\mu$m \citep[see also Section 
\ref{intro}, and][]{Kozasa09,Szalai11}. Therefore the rapid grain destruction process seems to be less significant around these young SNe than it might be in older SNRs. 

We also found that single-temperature models underestimate 
the fluxes at 24 $\mu$m when there are MIPS data taken contemporaneously with the IRAC measurements (SNe 2003ie, 2006bp, 2006my). 
In these cases we added a cold component ($T \sim$100 K) to our models, similar to \citet{Kotak09} and \citet{Szalai11}, which resulted in a reasonably good fit to the whole
observed SED. The parameters of the cold BBs are collected in Table \ref{tab:hotpar}.

The calculated dust masses for all SNe and all epochs are also collected in Table \ref{tab:dustpar} and \ref{tab:dustpar2}. Masses were determined using the 
formula

\begin{equation}
M_d = 4 \pi R^2 \tau_{\nu} / 3 \kappa_{\nu}
\end{equation}

\noindent adopted from \citet{Lucy89} and \citet{Meikle07}. It should be noted that these masses from analytic models are lower 
limits \citep{Meikle07}, because of the implicit assumption that the dust cloud has the lowest possible optical depth. As mentioned before, more realistic dust masses can 
be obtained only from numerical models for the optically thick dust cloud.

\begin{table}
\begin{center}
\caption{\label{tab:hotpar} Parameters of the additional cold and hot blackbodies.} 
\newcommand\T{\rule{0pt}{3.1ex}}
\newcommand\B{\rule[-1.7ex]{0pt}{0pt}}
\begin{tabular}{lcccc}
\hline
\hline
Object & Epoch & $T$ & $R$ & $L$\T \\
 & (days) & (K) & (10$^{16}$ cm) & (10$^{39}$ erg s$^{-1}$)\B \\
\hline
\multicolumn{5}{c}{Cold component}\T \\
\hline
SN~2003ie & 488 & 90 & 29.5 & 11.6\T \\
SN~2006bp & 401 & 110 & 24.0 & 6.0 \\
 & 767 & 80 & 42.0 & 5.1 \\
SN~2006my & 342 & 120 & 33.5 & 18.0 \\
 & 559 & 90 & 76.0 & 28.6 \\
\hline
\multicolumn{5}{c}{Hot component}\T \\
\hline
SN~2004A & 247 & 4800 & 0.045 & 76.1\T \\ 
SN~2006my & 205 & 3700 & 0.105 & 146.3\B \\
\hline
\end{tabular}
\end{center}
\end{table}

To check this statement and the trustworthiness of our analytic model fitting, we generated some models with the three-dimensional radiative-transfer code MOCASSIN 2.02.55. 
\citep{Ercolano03, Ercolano05}. This numerical code uses a ray-tracing technique, following the paths of photons emitted from a given source through a spherical shell 
containing a specified medium. The shell is mapped onto a Cartesian grid allowing light-matter interactions (absorption, re-emission, and scattering events) and to track 
energy packets until they leave the shell and contribute to the observed SED. It is applicable for reconstructing dust-enriched environments of CC SNe and to 
determine physical parameters such as grain-size distribution, composition, and mass of dust \citep[see][]{Sugerman06,Ercolano07,Fabbri11,Szalai11}. The code handles 
different types of dust heating processes caused not only by photons but also by collisions of grains and electrons (B. Ercolano, personal comm.).

\begin{table*}
\begin{center}
\caption{\label{tab:numpar} Parameters for the best-fit MOCASSIN models to SN~2005af SED at +576 days.}
\newcommand\T{\rule{0pt}{3.1ex}}
\newcommand\B{\rule[-1.7ex]{0pt}{0pt}}
\begin{tabular}{ccccccc}
\hline
\hline
Composition & $L_*$ & $T_*$ & $R_{in}$ & $R_{out}$ & $n_{dust}$ & $M_{dust}$ \T \\
of dust & (10$^{5}$ L$_{\odot}$) & (K) & (10$^{16}$ cm) & (10$^{16}$ cm) & 10$^{-6}$ cm$^{-3}$ & (10$^{-5}$ M$_{\odot}$) \B \\
\hline
AC & 3.4 & 7000 & - & 0.8 & 150$^{\dagger}$ & 18.0 \T \\ 
Si:AC (0.65:0.35) & 16.9 & 8000 & 0.2 & 1.0 & 0.02$^{\ddagger}$ & 11.0 \\
\hline
\end{tabular}
\tablefoot{
$T_*$ and $L_*$ denote the temperature and the luminosity of the central illuminating blackbody source;
the geometry of the dust cloud was assumed to be a spherical shell with inner and outer radii $R_{in}$ and $R_{out}$. \\
$^{\dagger}$ Homogeneous (constant-density) spatial distribution of grains \\
$^{\ddagger}$ $\rho \propto$ r$^{-7}$ density profile for spatial distribution of grains}
\end{center}
\end{table*}

\begin{figure}
\centering
\resizebox{\hsize}{!}{\includegraphics{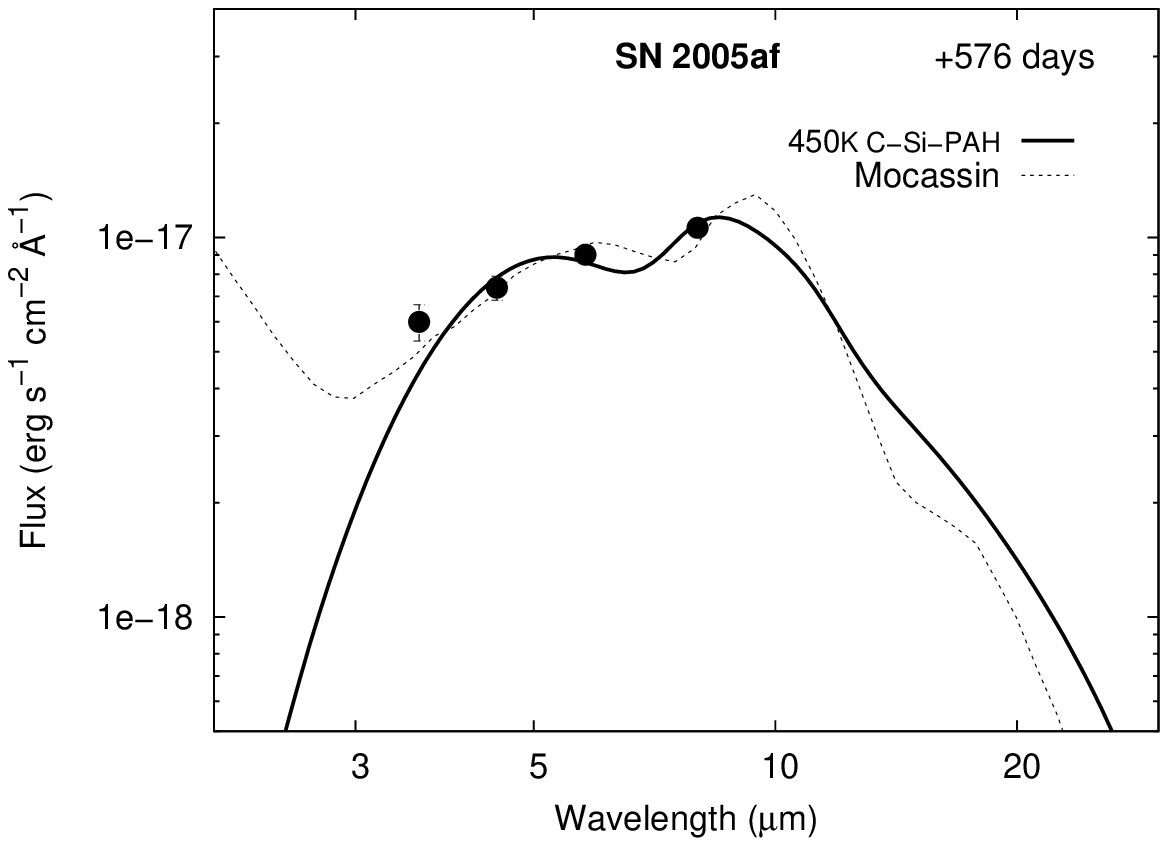}}
\caption{Best-fitting Si-containing analytical and numerical models for SN~2005af at +576 days (solid and dotted lines, respectively). The excess flux below 3 $\mu$m in 
the MOCASSIN model is due to the assumed hot central source, which was absent from the analytic models.}
\label{fig:2005af_analnum}
\end{figure}

We chose SN~2005af as a test object (at 576 days), for which we could fit both AC and Si-containing dust models. For AC grain composition we used the built-in optical 
constants from \citet{Hanner88}, while to fit also the flux at 8 $\mu$m we found a 0.65:0.35 Si-AC mixture to be the best solution \citep[built-in optical 
constants of astronomical silicate are taken from][]{Laor93}. We applied the MRN grain-size distribution.
The final results of numerical calculations are shown in Table \ref{tab:numpar}, while we compare the best-fitting analytical and numerical models in 
Figure \ref{fig:2005af_analnum}. The excess flux below 3 $\mu$m in the MOCASSIN model comes from the assumed hot central source, which was absent in the analytic 
models.

Although the two methods are very different (the analytic models describe the thermal MIR radiation of the dust, while the calculations of MOCASSIN are based on more detailed 
radiative transfer), we were able to generate similar SEDs in the MIR region with similar dust shell sizes. For pure AC dust, 
dust mass derived with the numerical model is an order of magnitude higher than that derived with the analytical model, 
which is very similar to our earlier results on SN~2004dj \citep{Szalai11}. Models for silicate dust are slightly 
different in numerical and in analytic models, and, to obtain the best fit, we had to apply larger grains ($a$ = 0.1-1 $\mu$m, MRN distribution). Despite these differences, 
the calculated mass for the silicate dust shell is very close to the value given by the analytic model.

\section{Discussion}\label{conc}

To understand the origin and heating mechanisms of dust in the environment of SNe, it is important to combine the MIR data with optical and/or 
near-IR measurements \citep[see e. g.][]{Kotak09,Fox10,Otsuka10,Szalai11,Meikle11}. Observations in X-ray and radio may also provide useful additional 
information about SN-CSM interactions. However, while {\it Spitzer} observations on the studied SNe were typically carried out in the nebular phase, there are 
only a very few additional late-time observations of these objects (see these cases above). Therefore, we can draw our conclusions based on mainly the MIR data analyzed 
in the previous Section. 

As mentioned before, there are three SNe in our sample (2003hn, 2005cs, and 2006bc) for which the available MIR data were not appropriate to resolve the point 
source at the position of the SN. The case of SN~2006bc is special, because there is a bright source (probably a 
surrounding \ion{H}{II} region) very close to the position of the SN. One would need high-resolution data and more sophisticated methods to extract information on this SN
\citep{Otsuka12}. Parallel to our studies, \citet{Gallagher12} carried out a complete optical and infrared data analysis of SN~2006bc based not only on {\it Spitzer} 
observations but also {\it HST} and {\it Gemini South} data. They found M$_{dust}$ $\leq$ 2 $\times$ 10$^{-3}$ M$_{\odot}$ newly formed dust in the ejecta, which is 
consistent with other similar cases.

The SEDs of the studied SNe and the derived parameters shown in Tables \ref{tab:bbpar}, \ref{tab:dustpar}, and \ref{tab:dustpar2} give guidelines and limits to the possible 
sources of the detected MIR radiation. We have the highest amount of data for SN~2005af, for which a sequence of MIR SEDs is available that shows a monotonically decreasing 
temperature (see Figure \ref{fig:sed}). This may be the sign of newly formed, cooling dust. It is also supported by the size of the dust sphere determined from the models:
the calculated expansion velocities are lower than 1300 km s$^{-1}$ in the nebular phase, which is low enough to be compatible with the inner ejecta velocities of 
II-P SNe. This is clearly another argument that most of the dust may have been formed in the ejecta. The calculated dust masses (1-3 $\times$ 10$^{-4}$ M$_{\odot}$) 
agree with the amount of SiO calculated by \citet{Kotak06} and with the results of \citet{Kotak08}, and they are also similar to the values found in other SNe.

\begin{figure}
\centering
\resizebox{\hsize}{!}{\includegraphics{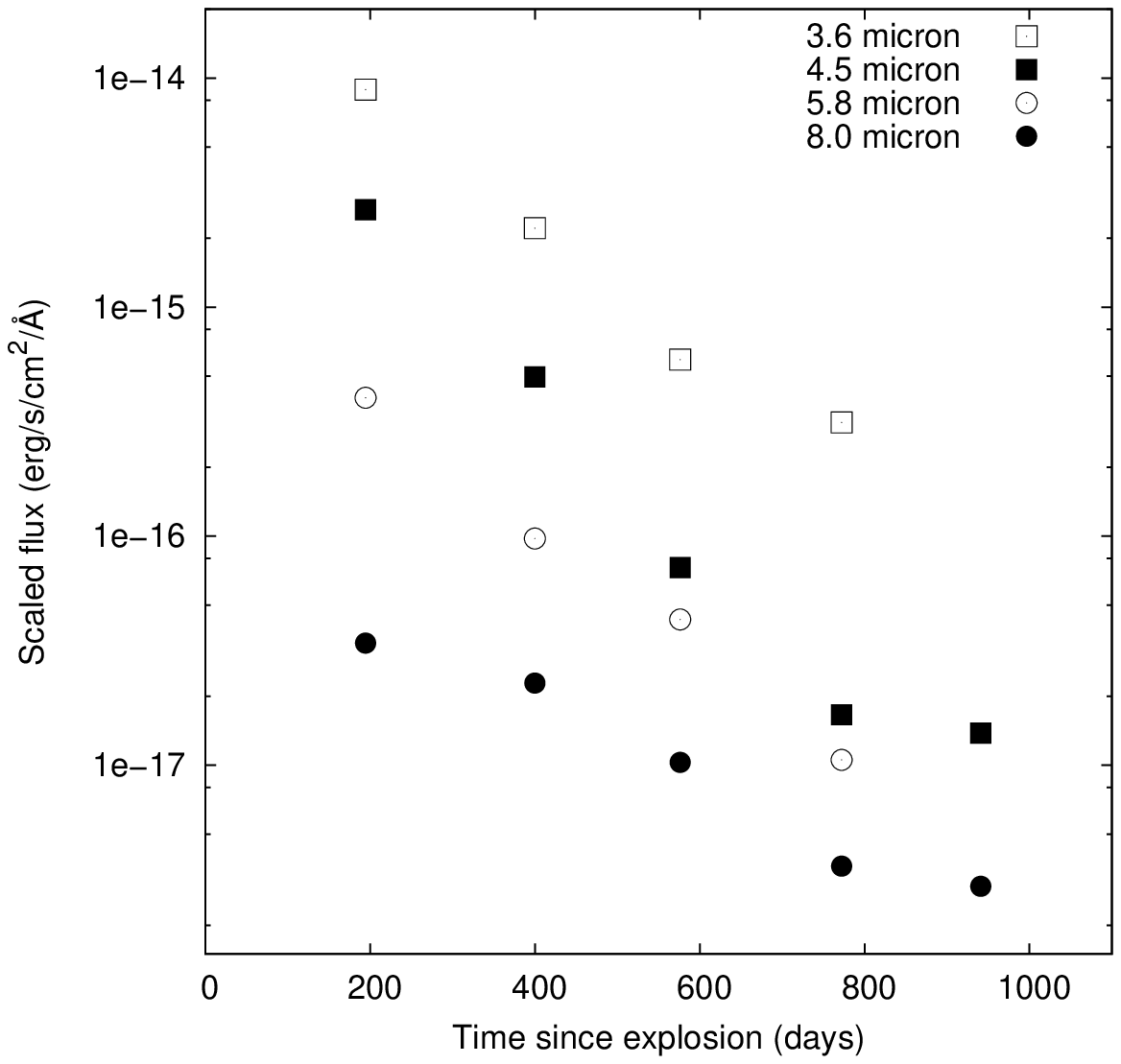}}
\caption{IRAC light curves of SN~2005af: 3.6 $\mu$m (open squares), 4.5 $\mu$m (filled squares), 5.4 $\mu$m (open circles), 8.0 $\mu$m (filled circles).}
\label{fig:2005af_lc}
\end{figure}

Another explanation for the observed behavior of MIR SEDs may be
the thermal radiation of pre-existing dust in the CSM re-heated by the SN, i. e., the IR-echo \citep{Bode80,Dwek83,Dwek85,Sugerman03}.   
We have very little information on the properties of the CSM around this SN. There are no published X-ray or radio measurements, and the relatively high 
value of the interstellar reddening ($E(B-V)_{gal}$ = 0.183 mag) is caused by the galactic ISM.
However, based on \citet{Dwek83} and \citet{Meikle06}, the shapes of the MIR light curves should be rather flat if there is an IR echo, but the IRAC light curves of 
SN~2005af decline monotonically in time (see Figure \ref{fig:2005af_lc}). Therefore we conclude that an IR echo is a less probable source of MIR radiation during the 
observed interval. It is also possible that dust grains condense in a cool dense shell (CDS) that is generated between the forward and reverse shock waves during 
the interaction of the SN ejecta and the pre-existing CSM. This is usually invoked for type IIn SNe, but it was also taken into account in a few other cases to 
explain the presence of early or very late MIR-excess \citep{Kotak09,Andrews10,Meikle11}. A CDS was also assumed to be the source of the cold dust component in the SEDs of II-P 
SNe \citep{Szalai11}.

The shapes of the MIR light curves and the presence of SiO make SN~2005af very similar to the well-studied SN~2004et. In this SN the main source of the warm component is also
identified as cooling, newly formed dust in the ejecta \citep{Kotak09}. Unfortunately, there are no {\it Spitzer} measurements on SN~2005af at very late epochs, so we could 
not see whether there is also a rise in IRAC fluxes after +1000 days like in SN~2004et \citep{Kotak09,Fabbri11}.

\begin{table*}
\caption{\label{tab:sn_spitzer} List of type II-P supernovae with published {\it Spitzer} data. SNe typed with italic were analyzed by our group (this is only partly true for 
SNe 2004dj and 2005af).}
\centering
\newcommand\T{\rule{0pt}{3.1ex}}
\newcommand\B{\rule[-1.7ex]{0pt}{0pt}}
\begin{tabular}{llllll}
\hline
\hline
Name & {\it Spitzer} data & Epochs & Detectable local dust & Dust mass & References \T \\
 & & & & (10$^{-5}$ M$_{\odot}$) & \B \\
\hline
SN~2002hh & IRAC, MIPS & 590,684,758 & Yes - IR echo & $\leq$ 3600 & 1, 2 \T \\
SN~2003gd & IRAC & 496-499,670-681,1264 & Yes - new, IR echo & 10 {\bf --} 2000  & 3, 4 \\
{\it SN~2004dj} & IRAC, MIPS, IRS & 98 -- 1381 & Yes - new, CDS,(maybe IR echo) & 1 {\bf --} 80 & 5, 6 \\
SN~2004et & IRAC, MIPS, IRS & 65 -- 1240 & Yes - new, IR echo & 4 {\bf --} 200 & 7, 8 \\
{\it SN~2005ad}$^{\dagger}$ & IRAC, IRS & 198, 215, 364 & Yes - new & 0.5 {\bf --} 1 & {\it this work} \\
{\it SN~2005af} & IRAC, MIPS, IRS & 194 -- 940 & Yes - new & 8 {\bf --} 21 & 9, 10, {\it this work} \\
SN~2006bc &  IRAC, MIPS & 537-540, 678-687, 753-775 & Yes - new & $\leq$ 200 & 11 \\
SN~2007it & IRAC & 351, 561, 718, 944 & Yes - new & 5 {\bf --} 70 & 12 \\
SN~2007od & IRAC, MIPS, IRS & 300, 455-470, 667 & Yes - new (maybe CDS) & 17 {\bf --} 19 & 13 \\
\\
{\it SN~2003J}$^{\dagger}$ & IRAC, MIPS & 471, 852, 2562 & Yes, partly & $\leq$ 710 & {\it this work} \\
{\it SN~2004A} & IRAC, IRS & 233-247, 406, 445, 563 & Yes, partly & $\leq$ 200 & {\it this work} \\
{\it SN~2007oc}$^{\dagger}$ & IRAC & 251, 415, 634, 759 & Yes, partly & $\leq$ 370 & {\it this work} \\
\\
{\it SN~2003ie} & IRAC, MIPS, IRS & 474-488, 517, 632, 1763 & No & -- & {\it this work} \\
{\it SN~2006bp} & IRAC, MIPS, IRS & 401-421, 628-671, & No & -- & {\it this work} \\
 & & 768-771, 1361 & & & \\
{\it SN~2006my} & IRAC, MIPS & 205, 342-354, 559-568 & No & -- & {\it this work} \\
{\it SN~2006ov} & IRAC, IRS & 321-343, 539 & No & -- & {\it this work} \B \\
\hline
\end{tabular}
\tablebib{
(1)~\citet{Barlow05}; (2) \citet{Meikle06}; (3) \citet{Sugerman06}; (4) \citet{Meikle07}; (5) \citet{Szalai11};\\ (6) \citet{Meikle11}; (7) \citet{Kotak09}; (8) 
\citet{Fabbri11}; (9) \citet{Kotak06}; (10) \citet{Kotak08}; (11) \citet{Gallagher12};\\ (12) \citet{Andrews11}; (13) \citet{Andrews10}.
}
\tablefoot{
$^{\dagger}$ From date of discovery.}
\end{table*}

There are fewer {\it Spitzer} data for the other eight SNe, and some of them were recorded during the warm mission (since the end of the cryogenic mission in 2009 only 
the IRAC 3.6 and 4.5 $\mu$m channels have been usable). Nevertheless, we may also draw some conclusions for these SNe. Among them we found obvious MIR flux 
changes in four objects: SNe 2003J, 2004A, 2005ad, and 2007oc (we note that for SN~2003J there is only one complete IRAC measurement, the second one was made 
during the warm mission). From these, SN~2005ad is the only object for which calculated parameters may indicate dust formation with high probability, because between the two 
observed epochs the 
dust model temperature decreased from 890 K to 750 K. The calculated expansion velocities are even lower than for SN 2005af, and the calculated 
dust masses ($\sim$10$^{-5}$ M$_{\odot}$) also agree with the values found in similar cases. Interestingly, for both SNe 2005af
and 2005ad, the calculated radii of the dust sphere are roughly constant (or decline slightly) in the nebular phase in spite of the expected expansion. \citet{Kotak09} found 
a similar effect in SN~2004et: their explanation was that the dust was contained within an optically thin cloud of optically thick dust clumps that cannot move outward 
because of the hotter, higher pressure interclump gas.

While the flux changes of SNe 2003J, 2004A, and 2007oc suggest that the observed MIR radiation is connected to the SN (at least partly), only a part of it may have
been emitted by newly formed grains in the ejecta. The derived dust sphere radii indicate very high expansion velocities that are much higher than the inner ejecta 
velocities of type II-P SNe in the nebular phase. At such large radii grain condensation is unlikely. 
SN~2004A shows changes in the dust temperature, and the calculated luminosities of the dust spheres are not much higher than for SNe 2005af and 2005ad, but 
the source of most of the MIR radiation may be pre-existing dust or the bulk of grains formed in a CDS. This may be even more valid for SNe 2003J and 2007oc.

For the remaining four SNe (2003ie, 2006bp, 2006my, 2006ov) there is no sign of real changes in the detected MIR fluxes. However, model fits to their 
SEDs suggest very high expansion velocities and much larger dust masses than for SNe 2005af and 2005ad (note that these values are still lower than the 
theoretically predicted dust masses). While this does not completely rule out that there is newly formed or pre-existing dust in the close environment of 
these SNe, the amount of dust should be smaller than the calculated values. 
We conclude that there are probably other sources of MIR radiation, e. g., dust and/or cold gas clouds within the 
host galaxies or in the line-of-sight of these SNe. For SN~2006bp, the high value of reddening, $E(B-V)_{total}$ = 0.4 mag, supports this idea. For the 
other three SNe the 
$E(B-V)$ values are much lower but are estimated only from the galactic reddening map of \citet{Schlegel98}. We also note that the 3$\arcsec$ aperture radii used 
by IRAC photometry correspond to 20-35 pc at the distances of our SNe, which is 2-3 orders of magnitude larger than the regions that the SN ejecta or shock waves 
could achieve at the given epochs (it may be less valid for SN~2005af, which is much closer than the other SNe in the sample).

In Table \ref{tab:sn_spitzer} we show all type II-P SNe that have {\it Spitzer} data analyzed and published to date including the objects presented in this work. 
As the final list shows, we doubled the number of objects, which might serve as a basis for statistical studies on a larger 
sample. We separated SNe that have some amount of local dust from those where there is no unambiguous sign for dust. Similar tables about the dust formation statistics of SNe 
can be found for instance in \citet{Fox11} and \citet{Gall11}.

\section{Summary}\label{summary}

We collected the public photometric and spectroscopic data of the {\it Spitzer Space Telescope} on 12 type II-P SNe. We found the data of nine of these objects 
to be appropriate for a detailed study, which almost doubles the number of type II-P SNe with a detailed, published MIR data analysis. 
We fit the observed SEDs with simplified one- or two-component dust models. In SNe 2005ad and
2005af we found cooling temperatures and decreasing luminosities of the warm component that are similar to the values found in other SNe that are thought to have 
newly formed dust in their environment.

The calculated temperatures for the other SNe do not show strong temporal variation, while the derived luminosities and radii are too high to be compatible with 
local dust. Moreover, the calculated dust masses in these cases are orders of magnitude higher than the observed amount of dust around the well-studied SNe listed above. The large 
radius of the warm component may suggest pre-existing dust in these cases, making it unclear if there was new dust formed around these SNe. Nevertheless, theoretical 
models predict orders of magnitude more newly formed dust in CC SNe. Our conclusions support the previous observational results that warm new dust in the environment of SNe 
contributes only slightly to cosmic dust content.

At the same time, our study may help to acquire better statistics on the dust formation capabilities of type II-P SNe, and present some objects that may be interesting for 
additional observations at longer wavelengths. In a few years the SNe for which we concluded the presence of some local dust (2003J, 2004A, 2005ad, 2005af, 2007oc) may show some 
other signs of dust formation in the transition phase.
 
\section*{Acknowledgments} 

We would like to express our thanks to the referee, A. P. Jones, for his valuable comments and suggestions, which helped us to improve the paper.
We thank B. Ercolano for sending us her MOCASSIN code ver. 2.02.55, and for her extensive help in running it. 
We also thank L. Colangeli and V. Mennella for providing the electronic version of their table about mass-extinction coefficients of carbon grains.
This work is supported by the Hungarian OTKA Grant K76816, by the European Union and is co-funded by
the European Social Fund through the T\'AMOP 4.2.2/B-10/1-2010-0012 grant.

\begin{appendix}
\section{Photometry}

\begin{center}
\begin{table*}
\footnotesize
\caption{\label{phot} Spitzer photometry for the clearly identifiable supernovae in our sample.}
\newcommand\T{\rule{0pt}{2ex}}
\newcommand\TT{\rule{0pt}{2ex}}
\newcommand\B{\rule[-0.7ex]{0pt}{0pt}}
\begin{tabular}{lcccccccc}
\hline\hline
~ &~ &~ & \multicolumn{6}{c}{Flux (10$^{-20}$ erg s$^{-1}$ cm$^{-2}$ \AA$^{-1}$)} \TT \B \\
\cline{4-9}
UT Date & MJD $-$ & $t - t_{expl}$ & \multicolumn{4}{c}{IRAC} & IRS PUI & MIPS \TT \\
~ & 2\,450\,000 & (days) & 3.6 $\mu$m & 4.5 $\mu$m & 5.8 $\mu$m & 8.0 $\mu$m & 13.0 - 18.5 $\mu$m & 24 $\mu$m \B \\
\hline
\hline
%
%
%
%
\multicolumn{9}{c}{SN~2003J} \T \B \\
\hline
2004-04-27$^{a}$ & 3123 & 471$^{\dagger}$ & 1069(126) & 706(67) & 1311(95) & 1684(121) & ... & ...\T \\
2005-05-13$^{a}$ & 3503 & 852$^{\dagger}$ & ... & ... & ... & ... & ... & 366(59) \\
2010-01-17$^{b}$ & 5214 & 2562$^{\dagger}$ & 750(116)) & 398(56) & ... & ... & ... & ... \B \\
\hline
\multicolumn{9}{c}{SN~2003ie} \T \B \\
\hline
2004-12-03$^{c}$ & 3342 & 474 & ... & ... & ... & ... & ... & 168(48) \T \\
2004-12-17$^{d}$ & 3356 & 488 & 152(44) & 77(23) & 363(36) & 515(38) & ... & ... \\
2005-05-10$^{d}$ & 3500 & 632 & 188(44) & 97(23) & 346(34) & 514(38) & ... & ... \\
2008-06-14$^{f}$ & 4631 & 1763 & 172(44) & 74(22) & 317(36) & 543(45) & ... & ... \B \\
\hline
\multicolumn{9}{c}{SN~2004A} \T \B \\
\hline
2004-09-10$^{d}$ & 3258 & 247 & 287(48) & 257(34) & 127(23) & 162(19) & ... & ... \T \\
2005-03-27$^{d}$ & 3456 & 445 & 122(34) & 68(19) & 70(19) & 123(18) & ... & ... \\
2005-07-23$^{e}$ & 3574 & 563 & 93(30) & 47(16) & 78(19) & 110(19) & ... & ... \B \\
\hline
\multicolumn{9}{c}{SN~2005ad} \T \B \\
\hline
2005-08-23$^{e}$ & 3605 & 198$^{\dagger}$ & 200(40) & 149(26) & 106(17) & 47(9) & ... & ... \T \\
2006-02-05$^{e}$ & 3771 & 364$^{\dagger}$ & 71(26) & 81(20) & 43(12) & 20(7) & ... & ... \B \\
\hline
\multicolumn{9}{c}{SN~2005af} \T \B \\
\hline
2005-07-22$^{e}$ & 3573 & 194 & 8912(243) & 26\,644(313) & 8039(135) & 3414(65) & ... & ... \T \\
2005-08-08$^{e}$ & 3590 & 211 &  & ... & ... & ... & 330(33) & ... \\
2006-02-12$^{e}$ & 3778 & 399 & 2213(121) & 4960(135) & 1955(67) & 2283(54) & ... & ... \\
2006-03-18$^{e}$ & 3812 & 433 & ... & ... & ... & ... & 260(26) & ... \\
2006-08-03$^{g}$ & 3950 & 571 & ... & ... & ... & ... & 340(34) & ... \\
2006-08-08$^{g}$ & 3955 & 576 & 590(66) & 729(53) & 867(47) & 1028(37) & ... & ... \\
2007-02-20$^{g}$ & 4151 & 772 & 314(51) & 166(27) & 211(26) & 363(26) & ... & ... \\
2007-03-18$^{g}$ & 4177 & 798 & ... & ... & ... & ... & 110(11) & ... \\
2007-07-16$^{h}$ & 4297 & 918 & ... & ... & ... & ... & ... & 23(13) \\
2007-07-24$^{h}$ & 4305 & 926 & ... & ... & ... & ... & ... & 28(14) \\
2007-08-07$^{h}$ & 4319 & 940 & ... & 138(26) & ... & 296(25) & ... & ... \B \\
\hline
\multicolumn{9}{c}{SN~2006bp} \T \B \\
\hline
2007-05-14$^{i}$ & 4234 & 401 & 355(53) & 257(33) & 464(36) & 618(37) & ... & ... \T \\
2007-06-03$^{i}$ & 4254 & 421 & ... & ... & ... & ... & ... & 119(15) \\
2007-12-19$^{j}$ & 4453 & 620 & ... & ... & ... & ... & 150(15) & ... \\
2007-12-27$^{j}$ & 4461 & 628 & 295(49) & 154(27) & 391(34) & 593(35) & ... & ... \\
2008-02-07$^{j}$ & 4504 & 671 & ... & ... & ... & ... & ... & 111(12) \\
2008-05-13$^{j}$ & 4600 & 767 & 285(48) & 160(27) & 445(35) & 585(36) & ... & ... \\
2008-05-17$^{j}$ & 4604 & 771 & ... & ... & ... & ... & ... & 107(16) \\
2008-06-05$^{j}$ & 4622 & 789 & ... & ... & ... & ... & 130(13) & ... \\
2009-12-29$^{b}$ & 5194 & 1361 & 283(49) & 140(26) & ... & ... & ... & ... \B \\
\hline
\multicolumn{9}{c}{SN~2006my} \T \B \\
\hline
2007-02-17$^{k}$ & 4148 & 205 & 967(91) & 528(50) & 542(41) & 803(37) & ... & ... \T \\
2007-07-03$^{j}$ & 4285 & 342 & 787(84) & 345(41) & 520(37) & 708(33) & ... & ... \\
2007-07-11$^{k}$ & 4292 & 349 & ... & ... & ... & ... & ... & 101(25) \\
2007-07-14$^{j}$ & 4295 & 352 & ... & ... & ... & ... & ... & 126(25) \\
2007-07-23$^{k}$ & 4304 & 361 & ... & ... & ... & ... & ... & 133(25) \\
2008-02-06$^{j}$ & 4502 & 559 & 731(82) & 312(40) & 504(37) & 742(35) & ... & ... \\
2008-02-15$^{j}$ & 4511 & 568 & ... & ... & ... & ... & ... & 126(35) \B \\
\hline
\multicolumn{9}{c}{SN~2006ov} \T \B \\
\hline
2004-06-10$^{a}$ & 3166 & -897 & 508(109) & 356(53) & 789(114) & 822(214) & ... & ... \T \\
2007-07-03$^{j}$ & 4284 & 321 & 606(108) & 415(53) & 752(112) & 1162(200) & ... & ... \\
2008-02-06$^{j}$ & 4502 & 539 & 567(110) & 363(53) & 780(118) & 815(221) & ... & ... \B \\
\hline
\multicolumn{9}{c}{SN~2007oc} \T \B \\
\hline
2008-07-12$^{l}$ & 4659 & 251$^{\dagger}$ & 477(62) & 510(48) & 297(30) & 388(27) & ... & ... \T \\
2008-12-23$^{l}$ & 4823 & 415$^{\dagger}$ & 287(49) & 192(31) & 147(22) & 214(20) & ... & ... \\
2009-07-30$^{m}$ & 5042 & 634$^{\dagger}$ & 140(36) & 22(12) & ... & ... & ... & ... \\
2009-12-02$^{m}$ & 5167 & 759$^{\dagger}$ & 174(39) & 46(16) & ... & ... & ... & ... \B \\
\hline
\hline
\end{tabular}
\tablefoot{
$^{a}$PID 69 Fazio et al.;
$^b$PID 61009 Freedman et al.;
$^c$PID 3124 Alexander et al.;
$^d$PID 3248 Meikle et al.;
$^e$PID 20256\\ Meikle et al.;
$^f$PID 484 Gorjian et al.;
$^g$PID 30292 Meikle et al.;
$^h$PID 40410 Rieke et al.;
$^i$PID 30496 Fisher et al.;\\
$^j$PID 40619 Kotak et al.;
$^k$PID 30945 Kenney et al.;
$^l$PID 50534 Andrews et al.;
$^m$PID 60071 Andrews et al.\\
$^{\dagger}$ From date of discovery.\\
}
\end{table*}
\end{center}

\end{appendix}
\end{document}